\documentclass[sigplan,10pt]{acmart}
\acmSubmissionID{123}
\renewcommand\footnotetextcopyrightpermission[1]{}
\setcopyright{none}
\usepackage{color}
\usepackage{appendix}
\usepackage{wrapfig}
\usepackage{multirow}
\usepackage{xurl}
\usepackage{algorithmicx,algorithm}
\usepackage[noend]{algpseudocode}
\usepackage[subrefformat=parens,labelformat=parens,caption=false]{subfig}
\usepackage{enumitem} 
\usepackage{times}
\usepackage{xspace}
\usepackage{xcolor}
\usepackage{latexsym}
\usepackage{amsmath}
\usepackage{amsthm}
\usepackage{booktabs}
\usepackage{tabularx}
\usepackage{balance}
\usepackage{graphicx}
\usepackage[normalem]{ulem}
\usepackage{svg}
\svgsetup{
    inkscapepath=i/svg-inkscape/
}
\svgpath{{svg/}}
\usepackage{etoolbox}
\usepackage[all]{nowidow}
\usepackage{listings} 
\usepackage{subcaption}

\usepackage{cleveref}
\usepackage{pifont}
\usepackage{makecell}
\usepackage{tikz}
\usetikzlibrary{positioning}
\usepackage{pgfplots}
\pgfplotsset{compat=1.18}

\crefformat{section}{\S#2#1#3}
\crefformat{subsection}{\S#2#1#3}
\crefformat{subsubsection}{\S#2#1#3}

\newcommand{\mysection}[1]{\vspace{-0em}\section{#1}}
\newcommand{\mysubsection}[1]{\vspace{-0em}\subsection{#1}}
\newcommand{\mysubsubsection}[1]{\vspace{-0em}\subsubsection{{\normalsize #1}}}

\setlist{nolistsep}
\lstdefinelanguage{json}{
  basicstyle=\ttfamily\tiny,
  showstringspaces=false,
  breaklines=true,
  string=[b]",
  morecomment=[l]{//},
  morecomment=[s]{/*}{*/},
}

\newcommand{\cut}[1]{}

\newcommand{\sysname}{Pythia\xspace}

\newcommand{\naive}[0]{na\"{i}ve\xspace}

\newcommand{\naively}[0]{na\"{i}vely\xspace}

\newcommand{\codeIn}[1]{{\small\texttt{#1}}}

\newcommand{\squishlist}{
   \begin{list}{$\bullet$}
    { \setlength{\itemsep}{0pt}      \setlength{\parsep}{3pt}
      \setlength{\topsep}{3pt}       \setlength{\partopsep}{0pt}
      \setlength{\leftmargin}{1.0em} \setlength{\labelwidth}{1em}
      \setlength{\labelsep}{0.5em} } }
\newcommand{\squishend}{
    \end{list}  }
    
\newtheorem{theorem}{Theorem}[section]

\newtheorem{definition}[theorem]{Definition}

\newtheorem{lemma}[theorem]{Lemma}

\newif\ifdraft
\drafttrue

\microtypecontext{spacing=nonfrench}

\definecolor{javared}{rgb}{0.6,0,0} 
\definecolor{javagreen}{rgb}{0.25,0.5,0.35} 
\definecolor{javapurple}{rgb}{0.5,0,0.35} 
\definecolor{javadocblue}{rgb}{0.25,0.35,0.75} 

\newcommand{\MyPara}[1]{\smallskip\noindent\textbf{#1}~}

\newcommand{\captionfonts}{\small}

\makeatletter  
\long\def\@makecaption#1#2{%
  \vskip\abovecaptionskip
  \sbox\@tempboxa{{\captionfonts #1: #2}}%
  \ifdim \wd\@tempboxa >\hsize
    {\captionfonts #1: #2\par}
  \else
    \hbox to\hsize{\hfil\box\@tempboxa\hfil}%
  \fi
  \vskip\belowcaptionskip}
\makeatother   

\newcommand{\squishlistree}{
   \begin{list}{$\bullet$}
    { \setlength{\itemsep}{0pt}      \setlength{\parsep}{0pt}
      \setlength{\topsep}{3pt}       \setlength{\partopsep}{0pt}
      \setlength{\leftmargin}{1em} \setlength{\labelwidth}{1em}
      \setlength{\labelsep}{0.5em} } }

\newcommand{\squishlisttwo}{
   \begin{list}{$\bullet$}
    { \setlength{\itemsep}{0pt}    \setlength{\parsep}{0pt}
      \setlength{\topsep}{0pt}     \setlength{\partopsep}{0pt}
      \setlength{\leftmargin}{2em} \setlength{\labelwidth}{1.5em}   
      \setlength{\labelsep}{0.5em} } }

\newcommand{\eg}{\hbox{\emph{e.g.}}\xspace}

\makeatletter
\newcommand*{\circled}{\@ifstar\circledstar\circlednostar}
\makeatother
 
\newcommand*\circledstar[1]{%
   \tikz[baseline=(C.base)]
     \node[%
       fill=black!20,
       circle,
       minimum size=1em,
       text=black,
       font=\footnotesize,
       inner sep=0.3pt
     ](C) {#1};%
}
\newcommand*\circlednostar[1]{%
   \tikz[baseline=(C.base) - .6em]
     \node[%
       fill=black,
       text=white,
       circle,
       minimum size=.8em,
       font={\bf \footnotesize},
       inner sep=0.2pt
     ](C) {#1};%
}

\begin{document}
\sloppy
\date{}
\fancyhead{}
\title{\sysname: Exploiting Workflow Predictability for Efficient Agent-Native LLM Serving}

\author{
Shan Yu$^{1}$\textsuperscript{*},\enspace
Junyi Shu$^{1}$\textsuperscript{*}\textsuperscript{$\dagger$},\enspace
Yuanjiang Ni$^{2}$,\enspace
Kun Qian$^{2}$,\enspace
Xue Li$^{3}$,\enspace
Yang Wang$^{4}$,\enspace
Jinyuan Zhang$^{1}$,\enspace \\
Ziyi Xu$^{5}$,\enspace
Shuo Yang$^{6}$,\enspace
Lingjun Zhu$^{3}$,\enspace
Ennan Zhai$^{2}$,\enspace
Qingda Lu$^{2}$,\enspace
Jiarong Xing$^{7}$,\enspace
Youyou Lu$^{8}$,\enspace \\
Xin Jin$^{9}$,\enspace
Xuanzhe Liu$^{9}$,\enspace
Harry Xu$^{1}$\\[0.5ex]
$^1$UCLA \quad
$^2$Alibaba Cloud Computing\quad
$^3$Alibaba Group \quad
$^4$Intel \quad
$^5$SJTU \quad \\
$^6$UC Berkeley \quad
$^7$Rice University \quad
$^8$Tsinghua University \quad
$^9$Peking University
}

\begin{abstract}

As LLM applications grow more complex, developers are increasingly adopting multi-agent architectures to decompose workflows into specialized, collaborative components, introducing structure that constrains agent behavior and exposes useful semantic predictability. Unlike traditional LLM serving, which operates under highly dynamic and uncertain conditions, this structured topology enables opportunities to reduce runtime uncertainty\textemdash yet existing systems fail to exploit it, treating agentic workloads as generic traffic and incurring significant inefficiencies. Our analysis of production traces from an agent-serving platform and an internal coding assistant reveals key bottlenecks, including low prefix cache hit rates, severe resource contention from long-context requests, and substantial queuing delays due to suboptimal scaling. To address these challenges, we propose \sysname, a multi-agent serving system that captures workflow semantics through a simple interface at the serving layer, unlocking new optimization opportunities and substantially improving throughput and job completion time over state-of-the-art baselines.
\renewcommand{\thefootnote}{\fnsymbol{footnote}}
\footnotetext[1]{Shan Yu and Junyi Shu contributed equally.}
\footnotetext[2]{Junyi Shu is the corresponding author.}
\end{abstract}
\maketitle
\mysection{Introduction}
\label{sec:introduction}

The conventional wisdom in general LLM serving is that \emph{everything is unpredictable}: neither the types nor the number of models activated at any given time can be anticipated, and the same holds for the types and volumes of requests directed to each model. This uncertainty stems from the fact that model providers typically expose only API endpoints for users to access their models, without control over when or how those models are invoked. Consequently, most prior work on LLM serving~\cite{vLLM:SOSP23,SGLang:NIPS24,Sarathi-Serve:OSDI24, Prism} assumes a highly dynamic execution environment, where nearly every configuration parameter of the serving system~(\eg, KV cache size, sharing mode, or autoscaling level) must be adjusted adaptively based on the workload observed at runtime. While such adaptability offers considerable flexibility, a fully reactive serving system often converges to suboptimal policies.

\MyPara{Production trace analysis.} To demonstrate how existing black-box approaches stifle efficiency, we analyzed large-scale production traces from our agent-serving service. We conducted in-depth profiling of an internal multi-agent coding assistant. Our analysis~(\S\ref{sec:motivation}) exposes three fundamental challenges in serving agentic workloads that contradict common assumptions. First, cache effectiveness is far lower than expected: specialized agents use distinct prompts and are invoked with long temporal gaps, causing previously cached prefixes to be evicted and leading to frequent zero-hit scenarios and long prefill latency. Second, the system suffers from severe resource inefficiency due to workflow-agnostic scheduling: heterogeneous requests are arbitrarily mixed, creating load imbalance across replicas and models, inducing memory contention, frequent preemptions, and costly recomputation. 
Third, multi-agent workflows exhibit sharp, structured bursts\textemdash often exceeding 50\% spikes within a minute\textemdash that overwhelm serving capacity; because existing systems rely on reactive, smoothed autoscaling without awareness of workflow dependencies, they fail to anticipate these cascades, resulting in queuing and congestion across multiple LLMs.

\begin{figure}[t]
  \centering
	\captionsetup[subfloat]{captionskip=0pt}
	\subfloat[Our internal coding assistant.]{
		\label{fig:workflow:coding}
		\includegraphics[width=0.9\linewidth]{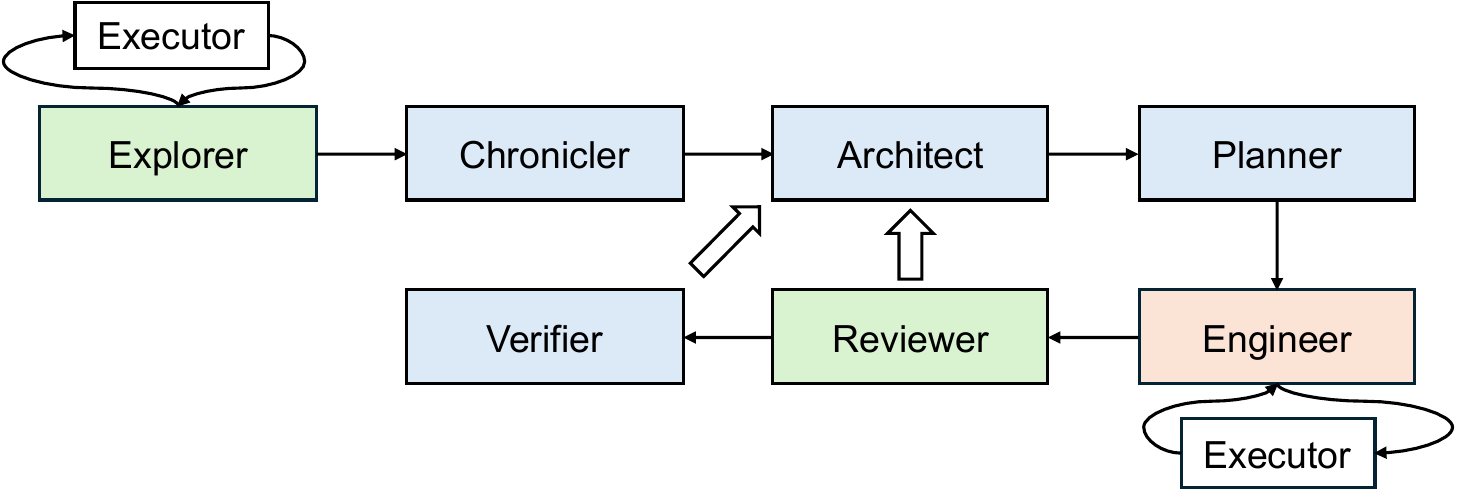}}
    \\
    \subfloat[An example of deep research agent.]{
		\label{fig:workflow:research}
		\includegraphics[width=0.9\linewidth]{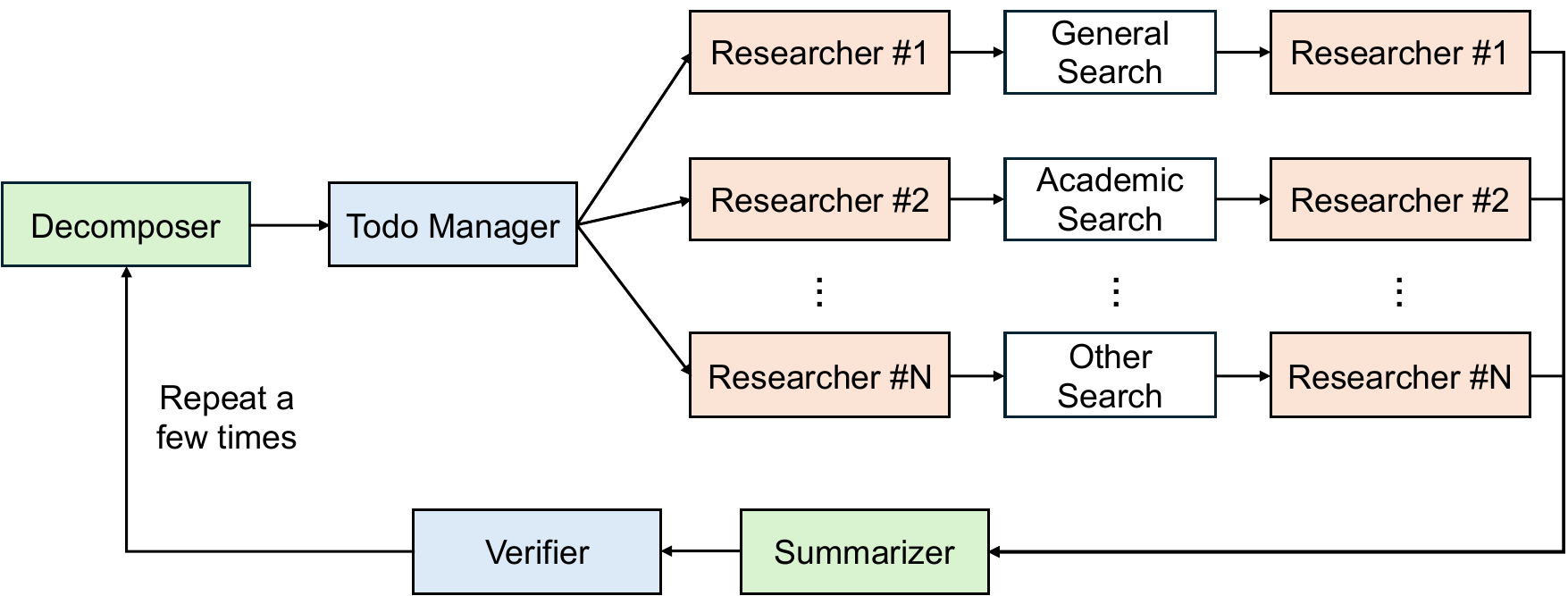}}
  \caption{Examples of multi-agent workflows. 
  \label{fig:workflow}}
\end{figure}

\MyPara{Key insight: Multi-agent workflows are predictable.} Unlike general LLM serving which is fundamentally unpredictable, multi-agent workflows (as illustrated in Figure~\ref{fig:workflow}) present a remarkably different scenario. While the underlying models might be general-purpose, the fixed set of agent types and their specific roles within the workflow tightly constrain how these models are invoked.

In such workflows, high-level reasoning components\textemdash often powerful LLMs acting as architects, planners, or decomposers\textemdash process complex task requirements and break them down into a set of sub-tasks. Each sub-task is then handed off to a specialized worker agent (\eg, an engineer or researcher) equipped with a particular~(often smaller) model and a set of tools.
Each agent typically performs a well-defined task and interacts with its model in a consistent manner. In many industrial deployments, a GPU cluster is dedicated to serving a small set of particular agentic applications (\eg, coding assistants, deep-research systems). 

As a result, requests from the same agent tend to follow similar usage patterns, introducing a substantial degree of \emph{predictability in resource usage}.

Resource predictability in a multi-agent application arises from two key factors: \emph{(1) predictable workflow} and \emph{(2) predictable workload}. First, at serving time, the structure of the workflow\textemdash often represented as a graph in which nodes correspond to individual agents and edges denote data flow\textemdash is either fully determined~(in static workflows) or follows repetitive patterns~(in workflows that evolve dynamically through reasoning). This structural information reveals how requests propagate across agents and enables more informed optimization. For example, a worker agent that processes entirely new inputs each time derives minimal benefit from prefix caching. Likewise, when agents are deployed across multiple GPUs, scheduling can be designed to rapidly saturate the pipeline and maximize utilization, rather than relying on a \naive FCFS policy that simply prioritizes earlier-arriving requests.

Second, since each worker~(non-orchestrator) agent is responsible for a specific task with a bounded level of complexity\textemdash where complex problems are typically decomposed into a collection of simpler subtasks\textemdash the output length and execution time of a request at each agent node typically follow relatively simple distributions, making them amenable to accurate estimation via dry runs or online profiling.
Once characterized, these estimates can be leveraged during workflow execution to enable precise ahead-of-time cache warm-up and proactive model autoscaling prior to invoking an agent.

\MyPara{\sysname.} Building on this insight, we designed \sysname, an agent-native serving system that systematically exploits workflow predictability to unlock optimizations previously considered intractable. Traditional serving engines treat agentic reasoning as an entirely opaque and dynamic process, limiting them to reactive, suboptimal resource allocation. In contrast, \sysname realizes proactive, global coordination by approximating a request's remaining execution time\textemdash inferred from its current position in the workflow graph and the expected output footprint of its specific agent role. By translating these structural insights into a priority mechanism that schedules requests based on anticipated completion times, \sysname enables fine-grained optimizations that fundamentally elude workflow-agnostic systems, significantly improving both end-to-end performance and cluster utilization.

\MyPara{Challenges.} 
Fully leveraging these aspects of predictability for effectively serving multi-agent applications requires overcoming the following two challenges. 

The first challenge is \emph{how to inform the serving system about what aspects of the workflow are predictable}. For example, the workflow graph is a critical piece of information that needs to be conveyed to the serving system. However, as agent development becomes increasingly accessible to a broad audience~(\eg, as illustrated by OpenClaw~\cite{openclaw}), many agent developers may not be expert programmers and therefore may not fully understand or be able to specify the underlying graph structure. This difficulty is further compounded by the fact that real-world multi-agent applications often generate dynamic graphs, where agent nodes are created by the orchestrator according to the complexity of the initial task. 

To solve the problem, \sysname introduces our core contribution: an \emph{example-driven synthesis} approach~\cite{synthesize-regular-expressions}, which automatically synthesizes a graph representation from the profiling data. This approach builds on two important observations. First, most agentic applications have straightforward structures and a few runs with representative inputs can already generate enough profiling data that can help us construct the workflow graph. Second, although a dynamic graph cannot be statically described, it often consists of repetitive patterns (\eg, consecutive Engineer-Executor interactions as in Figure~\ref{fig:workflow:coding}). By analyzing these patterns, the dynamic execution flow can be entirely expressed and predicted using \emph{regular expressions}.

To this end, \sysname only requires agentic frameworks~\cite{Autogen:COLM24,langgraph, langchain, openclaw} to pass lightweight information about application 

and agent types via the \codeIn{extra\_body} payload already supported by OpenAI-compatible endpoints~\cite{openai-api} without making any code change in user applications. At the gateway, \sysname employs a profiler to derive workflow and workload insights\textemdash including a regular expression capturing the workflow graph, as well as per-agent output length\textemdash and injects these annotations back into the payload for downstream consumption by the serving system.

The second challenge is \emph{how to take advantage of such predictive information in the serving system.} Traditional LLM serving engines rely on a reactive paradigm that manages memory and schedules requests based on the current system state. Naively transitioning to a proactive model introduces the risk of severe performance penalties; aggressive, look-ahead decisions can easily trigger destructive resource thrashing and premature cache evictions if not carefully managed. To mitigate these risks and safely realize these optimizations, 
\sysname introduces three coordinated mechanisms\textemdash each aligned with the key opportunities (\S\ref{sec:motivation}) identified in our trace analysis\textemdash to proactively optimize execution. At the node level, a \emph{speculative cache manager} enhances reactive LRU with a Belady-inspired policy, leveraging workflow insights to prefetch shared contexts, evict transient tokens, and asynchronously warm caches to mask prefill latency. At the cluster level, a \emph{lookahead request scheduler} uses predicted output lengths and workflow structure to balance heterogeneous workloads and prioritize execution, thereby reducing memory contention and head-of-line~(HoL) blocking. Complementing these components, a \emph{phase-adaptive autoscaler} anticipates structural workload shifts, such as fan-outs and phase transitions, to proactively scale models up or down, mitigating burst-induced queuing and improving overall resource efficiency.

\MyPara{Results.} We have implemented \sysname on top of SGLang and evaluated it across representative multi-agent applications. Our results demonstrate up to 2.9$\times$ average JCT reduction with 1.96$\times$ throughput improvement. \sysname is currently in the phase of extensive testing and pilot deployment at a larger cluster scale internally for broader adoption.

\begin{figure}
  \centering
  \includegraphics[width=0.6\linewidth]{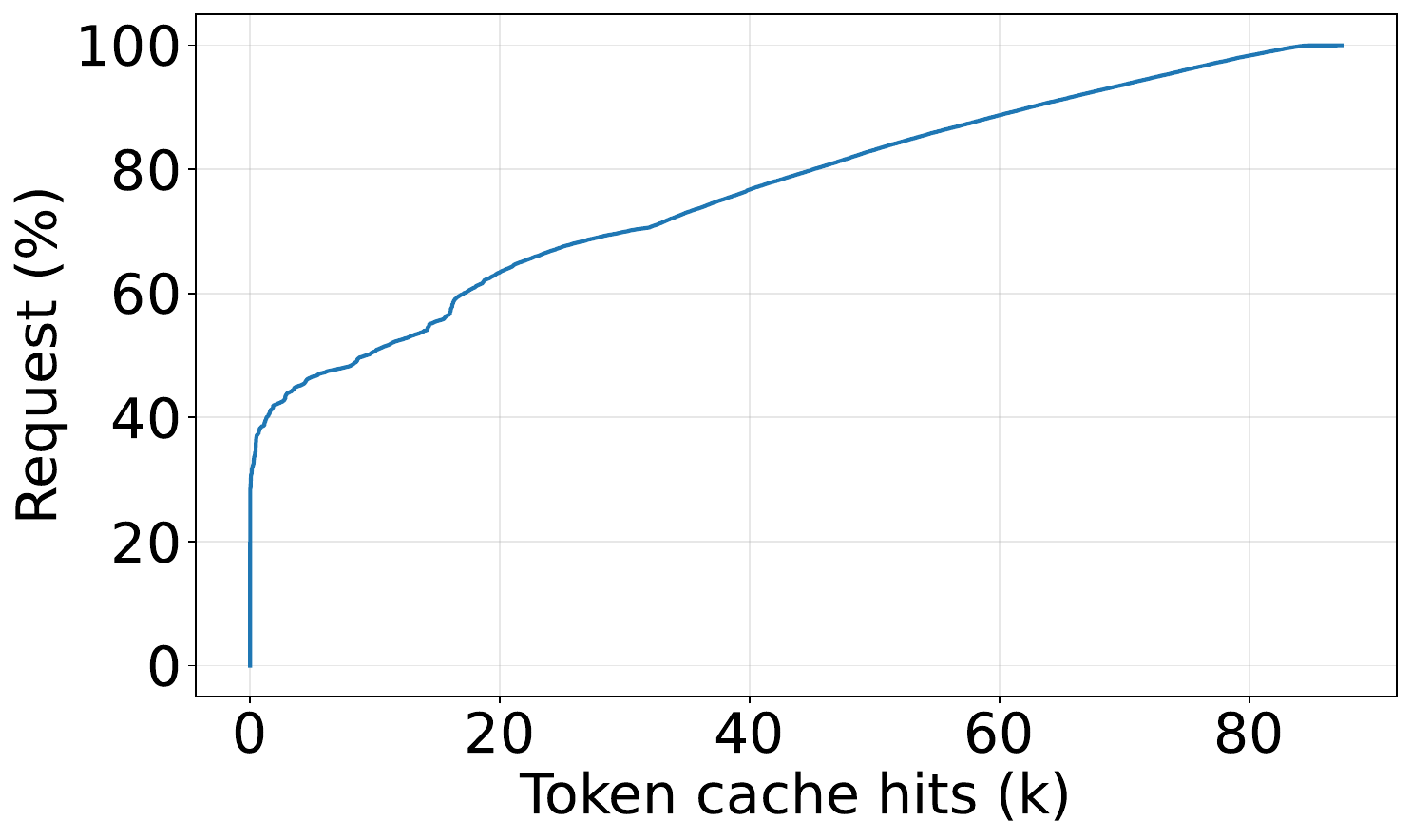}
  \caption{CDF depicting the percentages of requests with various numbers of token hits from our agent-serving platform.} 
  \label{fig:cache-hit}
\end{figure}

\mysection{Background and Motivation}
\label{sec:motivation}

\subsection{Background}

\MyPara{LLM serving.} Today's LLM serving systems are architected around request-level and token-level execution~\cite{vLLM:SOSP23, SGLang:NIPS24, Orca:OSDI22, DistServe:OSDI24, Sarathi-Serve:OSDI24}. Performance optimizations heavily target isolated micro-benchmarks, focusing on reducing Time-To-First-Token~(TTFT) and Time-Between-Tokens~(TBT) to accelerate prefill and decode phases. Crucially, the serving layer is usually assumed to have no prior knowledge of the output of ongoing requests and the incoming workload. Treating each request as an opaque, isolated entity forces the serving layer to rely on conservative, reactive policies. For example, request scheduling is typically relegated to FCFS queues, while prefix cache management relies strictly on reactive, recency-based eviction strategies.

\MyPara{Multi-agent workflow.} The interaction paradigm for LLMs is rapidly evolving from isolated, human-in-the-loop chat sessions to autonomous, goal-driven agents~\cite{openclaw, claude-yolo, codex-yolo}. Driven by this shift, the share of inference requests generated by agents executing multi-step workflows has surged over the past six months. At our agent-serving platform, over \textbf{80\%} of current requests are originated from agentic AI applications today.
In a modern multi-agent system, high-level objectives are decomposed into actionable plans, with specific subtasks delegated to specialized worker LLMs and tools. These worker agents\textemdash driven by tailored system prompts and often backed by domain-specific models\textemdash handle distinct roles such as planning, coding, and reviewing. Traditional serving systems treat agent requests as normal LLM requests, handling them in a way that is agnostic to agent semantics.

\MyPara{Hierarchical prefix cache.} Prefix caching stores computed key-value~(KV) tensors to reduce redundant prefills. Modern systems balance latency and capacity using a three-tier hierarchy~\cite{Mooncake:FAST25, LMCache}: L1 resides in scarce GPU HBM for immediate execution access; L2 acts as a spacious staging buffer in CPU host DRAM; and L3 utilizes shared storage to persist cross-engine contexts and minimize global cache misses.

\mysubsection{Challenges in Serving LLMs for a Multi-Agent Workflow}

As agentic AI increasingly dominates production workloads, MaaS providers including us, have introduced dedicated products with specialized billing models to support these applications~\cite{ali_coding_plan, byte_coding_plan, minimax_token_plan, glm_coding_plan}. Yet, a fundamental architectural mismatch persists: these services still rely on legacy, application-agnostic inference APIs. By flattening complex workflows into a series of opaque, independent requests, the serving layer is left entirely without knowledge of structural dependencies or distinct agent roles. To demonstrate how this black-box interface stifles efficiency, we analyzed large-scale production traces from our agent-serving service and conducted in-depth profiling of an internal multi-agent coding assistant deployed atop the service. This analysis reveals three major challenges where the lack of workflow visibility directly degrades system performance.

\begin{figure}[t]
  \centering
  \includegraphics[width=0.99\linewidth]{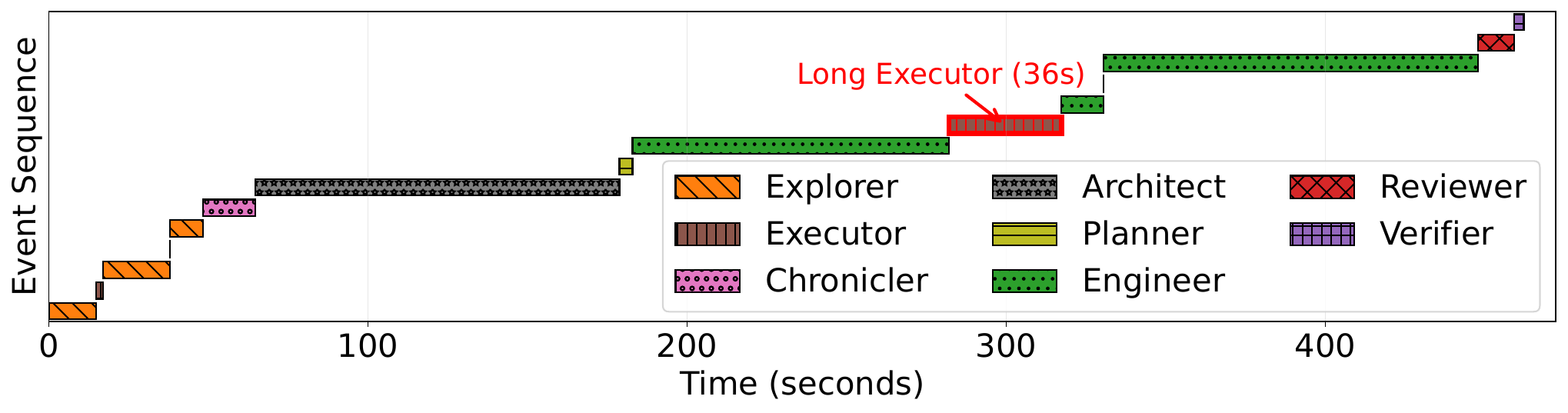}
  \caption{Timeline of the coding agent workflow: each bar represents an agent and  ``executor'' represents command line execution. 
  \label{fig:workflow-timeline}}
\end{figure}

\MyPara{Challenge \#1: Long prefill times due to unexpectedly low cache hit rates.} While existing work often assumes that agentic AI applications naturally achieve high cache hit rates due to repeated prompt structures~\cite{DualPath, ThunderAgent, Continuum}, our production traces reveal a starkly contradictory reality. Figure~\ref{fig:cache-hit} reports the distribution of the requests with different numbers of token cache hits. As shown, more than 40\% of requests have zero or very limited cache hit rates. This occurs because specialized agents rely on distinct system prompts, preventing prefix sharing across agent boundaries. 
Furthermore, the sequential execution of these workflows actively degrades cache retention. To illustrate this, Figure~\ref{fig:workflow-timeline} details the execution timeline of the coding assistant. We make two observations: (1) only a few consecutive LLM requests can benefit from prefix cache; and (2) there can be long-running tool calls~(\eg, code compilation) which can take tens of seconds between these requests. By the time an agent is re-invoked, its previously cached prefixes have mostly been flushed by the intervening traffic. 

\begin{figure}[t]
  \centering
  \begin{minipage}[t]{0.44\columnwidth}
    \centering
    \includegraphics[width=\linewidth]{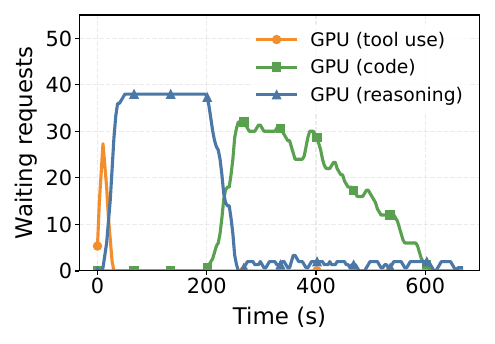}

    \small (a) Traffic imbalance between nodes in a workflow.
  \end{minipage}\hfill
  \begin{minipage}[t]{0.55\columnwidth}
    \centering
    \includegraphics[width=\linewidth]{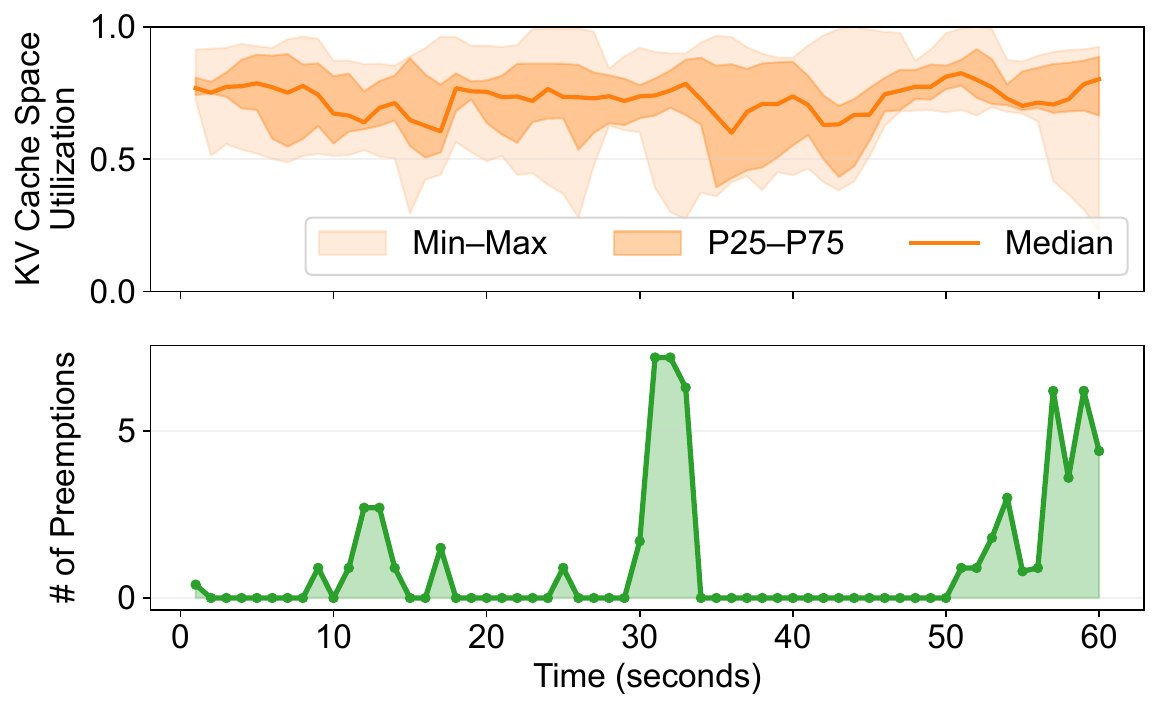}

    \small (b) Memory imbalance between nodes serving the same model.
  \end{minipage}

  \caption{Load imbalance and preemption when serving a multi-agent coding assistant.}
  \label{fig:lb-and-preempt}
\end{figure}

\MyPara{Challenge \#2: Wasted resources due to load imbalance and uncoordinated resource competition.} Figure~\ref{fig:lb-and-preempt} shows a one-minute serving trace of coding agents on an 8-GPU cluster, revealing a system heavily constrained by memory pressure and frequent request preemptions. This inefficiency arises from mixing heterogeneous requests\textemdash ranging from lightweight Planner responses to long, thousand-token Architect outputs\textemdash  without any awareness of the underlying workflow. We identify three key sources of this imbalance. First, \naive routing leads to significant load skew across data-parallel replicas of the same model. Second, resource allocation across models is uneven, leaving some saturated while others remain underutilized. Third, the lack of workflow awareness causes co-located agents to compete arbitrarily for shared GPU memory instead of being coordinated. As a result, overloaded replicas resort to preempting requests; without knowledge of expected output lengths, the system often evicts newer requests indiscriminately. These preempted requests must be recomputed from scratch, resulting in substantial computational waste and pronounced head-of-line blocking.

\begin{figure}[t]
  \centering
  \includegraphics[width=0.99\linewidth]{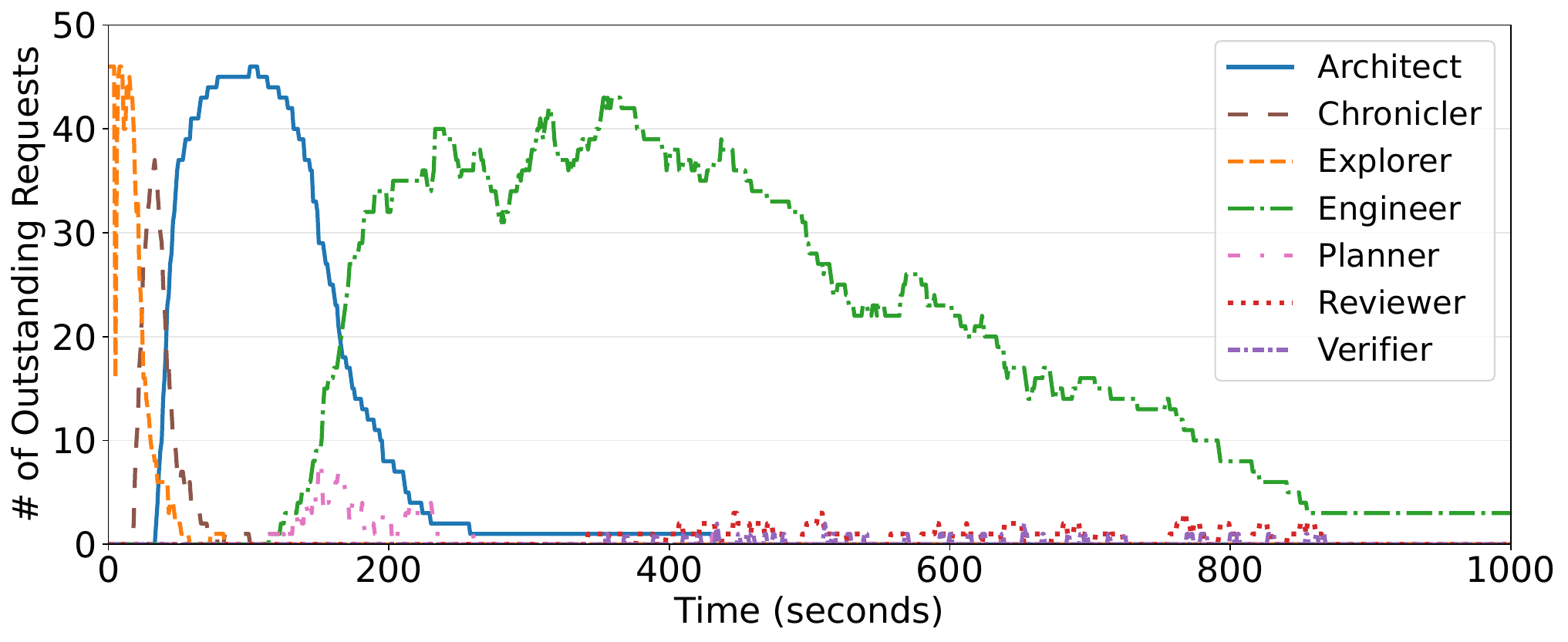}
  \caption{Outstanding requests of the multi-agent coding assistant for a batch of cron jobs. 
  \label{fig:burst}}
\end{figure}

\MyPara{Challenge \#3: Burst-induced queuing and congestion across LLMs.} Multi-agent workflows inherently exhibit extreme structural volatility; in our agent-serving platform, we observe that scheduled automated tasks~(\eg, batched cron jobs) 
can trigger request spikes of up to \textbf{50.3\%} for certain models within a single minute. As illustrated in Figure~\ref{fig:burst}, these massive surges rapidly overwhelm the multiple LLMs backing the system. 
Crucially, the interdependent nature of multi-agent workflows fundamentally exacerbates this issue. A burst at an entry node~(\eg, an Explorer agent) does not remain isolated; instead, it triggers a cascading wave of requests that rapidly propagates through downstream agents~(\eg, Chronicler, Architect, Engineer). However, because existing serving systems rely on smoothed, long-term historical metrics and workflow-agnostic scaling, they cannot anticipate this inter-model propagation. Standard reactive autoscaling requires tens of seconds to add capacity\textemdash a delay during which the burst has often already shifted to the next model in the pipeline, leaving the system perpetually lagging behind the wavefront and causing queue depths to surge.

\mysubsection{Clairvoyance in Agent-Native Serving}
\label{subsec:predictability}

To address the challenges outlined above, we must first understand why today's inference services fail to mitigate them. Current LLM serving engines are designed around an implicit assumption: requests are generated by independent human users. 
Human chat traffic is characterized by uncorrelated intents, unpredictable arrival times, and arbitrary prompt contents. Serving systems are forced to operate as strict online algorithms\textemdash managing caches,  scheduling requests, and scaling capacity reactively without any knowledge of the future.

However, agent traffic is generated by programs. Multi-agent workflows follow certain control flows, reuse established prompt templates, and exhibit access patterns dictated by their underlying source code. While the exact execution paths may be dynamically generated, the space of likely executions is far more constrained than human interactions. This programmatic nature implies a fundamental shift: if the serving system can capture the workflow's knowledge \emph{a priori}, it can transition from reactive, online guesswork to proactive, near-offline optimization. Specifically, we identify three critical dimensions of such knowledge that can systematically mitigate the observed bottlenecks.

\MyPara{Opportunity \#1: Workflow graph.} Capturing the structural dependency graph of a multi-agent application unlocks multiple critical system optimizations. First, it identifies recurring agent roles, allowing the serving layer to strategically retain their reusable prefixes rather than \naively evicting them under temporary memory pressure. Second, by understanding a request's approximate position within a workflow, the scheduler can estimate remaining execution times and dependencies, enabling critical-path-aware request scheduling. Finally, this structural awareness allows the system to proactively mitigate burst-induced queuing by anticipating imminent phase changes and fan-out patterns.

Acquiring this structural clairvoyance is highly feasible in production environments. While frameworks like AutoGen and LangGraph theoretically support fully dynamic, unconstrained agent group chats, real-world deployments rarely operate this way. 
Because unconstrained LLMs are susceptible to hallucination and task drift, enterprise applications typically impose stricter routing constraints and operational guardrails. For instance, our internal coding assistant enforces execution through hardcoded transition rules, using an LLM-based group chat selector only as a fallback for edge cases. As a result, production workflows tend to be stable, following highly probable execution paths with well-constrained iterative loops, leading to strongly concentrated transition probabilities between agent roles.

\MyPara{Opportunity \#2: Output length of ongoing requests.} Predicting the output length of active requests provides two critical operational advantages. First, it enables resource-aware scheduling to mitigate severe memory contention and load imbalance. By balancing the system based on predicted resource consumption rather than a \naive request count, the serving layer avoids the localized memory pressure that triggers catastrophic HoL blocking. Second, estimating the decode duration of an ongoing request provides a fine-grained timeline for when the request will arrive at the next agent in the workflow. This temporal awareness allows the serving system to fetch and warm the KV cache for the subsequent agents \emph{just in time}.

\begin{table}[t]
\centering
\caption{Output length statistics by agent role in the coding assistant. 
}
\label{tab:output_length}
\small
\begin{tabular}{lcc}
\toprule
Agent Role & Avg. Output Length & CV \\
\midrule
Explorer & 1924 & 0.45 \\
Chronicler & 912 & 0.26 \\
Architect & 1194 & 0.22 \\
Planner & 60 & 0.15 \\
Engineer & 3152 & 0.45 \\
Reviewer & 2620 & 0.18 \\
Verifier & 65 & 0.16 \\
\bottomrule
\end{tabular}
\end{table}

The basis for this predictability lies in the strict division of labor across the workflow. Because individual agents are scoped to specific, narrow tasks, their generative behaviors naturally converge into stable, recognizable profiles. As detailed in Table~\ref{tab:output_length}, Chronicler, Planner, Reviewer, and Verifier reliably produce relatively consistent lengths with a low coefficient of variation~(CV). Furthermore, while Explorer and Engineer's output lengths have higher variance due to their interactive nature, we observe a strong correlation between consecutive calls. Therefore, by observing an agent's semantic role and current context, the system can roughly bound a request's compute/memory footprint.

\MyPara{Opportunity \#3: The composition and lineage of future prompts.} Predicting the structure of upcoming prompts allows the serving system to proactively hide expensive prefill latency. Because specialized agents rely on distinct system prompts, they cannot natively share prefix caches; when a workflow transitions from one agent to another, the new request typically triggers a full, costly prefill computation. However, if the system predicts the next prompt, it can proactively compute and prepare the KV cache for the subsequent agent early, well before the actual request is issued. This proactive ``pre-prefill'' effectively eliminates the prefill bottleneck from the critical path.

The mechanics of multi-agent frameworks naturally expose the exact contents of these upcoming requests. A subsequent prompt is never a black box; it is reliably constructed by \emph{concatenating known static content~(\eg, the agent's specific system instructions) with the explicit inputs and generative outputs of the dependent agents that executed earlier in the workflow}.
Since the serving layer already tracks the workflow graph and holds the outputs from preceding steps, it possesses every ingredient required to accurately assemble and pre-compute the next prompt ahead of time.

\MyPara{Takeaway.} The programmatic nature of multi-agent workloads introduces strong structural, temporal, and semantic predictability. However, this valuable \emph{a priori} information is entirely lost across today’s stateless application–system interface. To address the resulting cascade of issues—cache thrashing, load imbalance, and burst-driven queuing, the serving layer must be endowed with a form of clairvoyance.

\mysection{\sysname Overview}
\label{sec:overview}

\begin{figure}[t]
    \centering
    \includegraphics[width=0.99\linewidth]{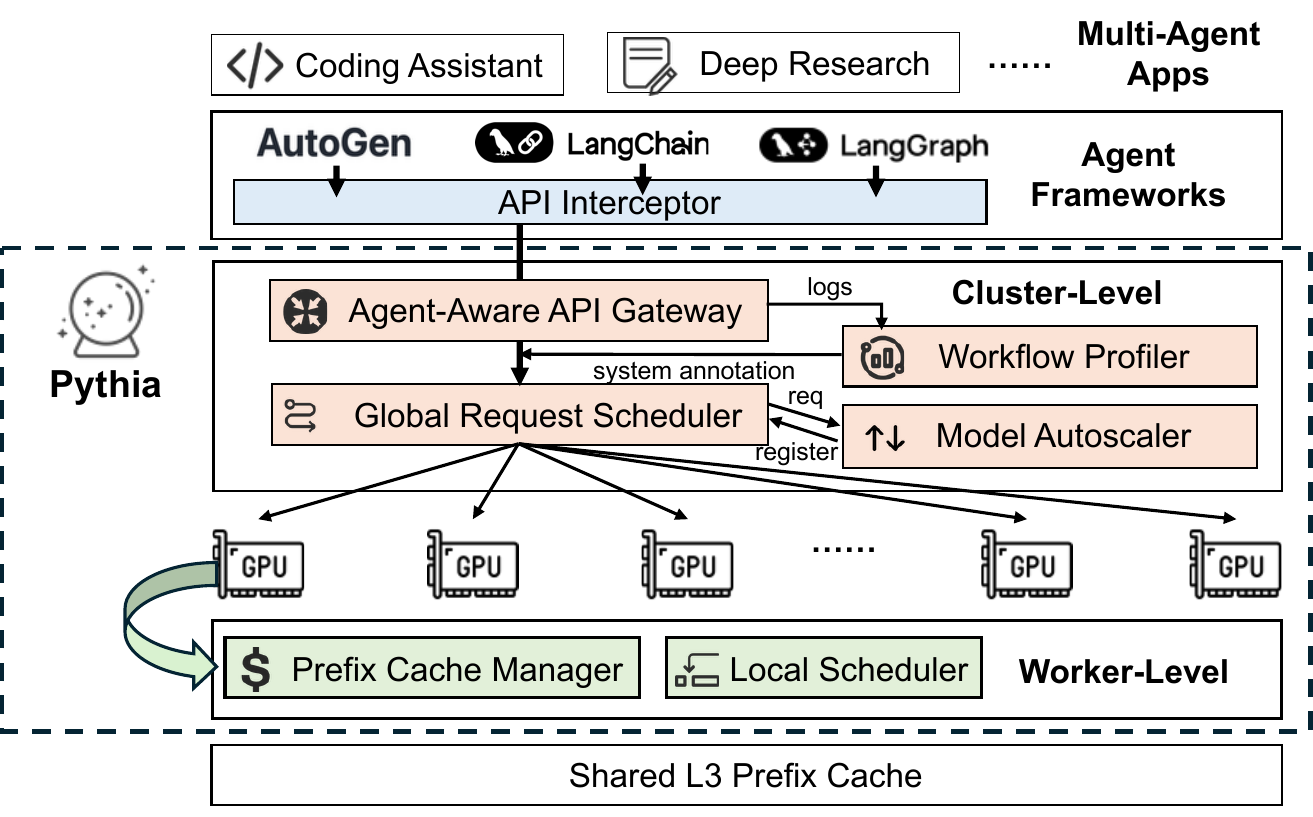}
    \caption{\sysname overview.}
    \label{fig:overview}
\end{figure}

Figure~\ref{fig:overview} depicts the system overview for \sysname.

\MyPara{Predictive information generation.}   
The entry point to \sysname is an OpenAI-compatible API (\S\ref{sec:design:predict}) extended to support lightweight metadata annotations. These simple annotations serve to identify distinct workflows and specific agent roles directly within popular frameworks~(\eg, LangGraph or AutoGen). This minimal API extension is completely transparent to end users and requires no changes to the application logic.

In addition to collecting the structural information of the multi-agent workflow, the profiler (\S\ref{sec:design:predict}) acts as the analytical core of \sysname by operating asynchronously outside the critical path. By continuously ingesting annotated request/response logs, it extracts and maintains the overarching control flow graphs and program characteristics of different agents.  Upon a new request, it uses the provided metadata to query these historical profiles, further annotating the requests with three critical execution estimates to downstream modules.

\MyPara{Predictive information consumption.}
\sysname uses the aforementioned predictive information at three distinct locations of the serving pipeline: per-node (agent) prefix cache management, global request scheduling, as well as per-node model scaling, which, respectively, correspond to the three major challenges faced by existing techniques (\S\ref{sec:motivation}).

Operating at the node level, the cache manager (\S\ref{sec:design:cache}) mitigates prefix thrashing by shifting from reactive LRU eviction to a proactive Belady-like strategy. It leverages the extracted workflow information to accurately prefetch shared contexts and immediately drop transient tokens that are no longer needed. Furthermore, by anticipating the precise prompt composition of upcoming requests, it asynchronously warms the cache for the next probable agent, effectively hiding prefill latency before the subsequent request even arrives.

Operating at the cluster level, the request scheduler (\S\ref{sec:design:schedule}) leverages the extracted program characteristics to execute resource-aware load balancing and precise request prioritization. By utilizing predicted output lengths, it evenly distributes heterogeneous workloads across data-parallel replicas. Furthermore, this profile-aware approach establishes better execution priorities, effectively mitigating localized memory contention and HoL blocking that drive request preemptions.

The per-node model autoscaler (\S\ref{sec:design:scale}) manages cluster-wide capacity by anticipating load changes rather than reacting to them. By evaluating the state of current active requests against the extracted workflow patterns, it forecasts imminent structural shifts, such as fan-outs and phase changes. This clairvoyance allows the system to proactively scale specialized models up before a burst hits, and scale them down when they are no longer needed, effectively alleviating burst-induced queuing.

We discuss how \sysname supports many distinct workflows, respond to pattern shifts, and handle adversarial behaviors in \S\ref{sec:discuss}.

\mysection{\sysname Design}
\label{sec:design}

\mysubsection{Producing Predictive Information}
\label{sec:design:predict}

To predict resource usage across serving stages, \sysname first extracts the semantic context of incoming requests. Our goal is to do so without imposing heavy integration overhead on application developers or violating the separation between application logic and serving infrastructure. To this end, we introduce a minimal, transparent API extension, coupled with an asynchronous background profiler that models workflow executions using classic program analysis techniques.

\MyPara{API abstraction.} 
Standard inference APIs isolate requests, stripping away the overarching application context. To bridge this gap, \sysname leverages the \codeIn{extra\_body} parameter natively supported by OpenAI-compatible endpoints. When an agentic framework dispatches a request, it injects a lightweight \codeIn{app\_metadata} payload identifying the execution context. As shown in Listing~\ref{lst:api_extension}, the framework only needs to supply three explicit identifiers:
\begin{enumerate}[leftmargin=1.5em,itemsep=1pt,parsep=1pt]
    \item \codeIn{workflow\_type\_id} uniquely identifies an application (\eg, \codeIn{coding\_assistant}), allowing the system to map the request to a specific structural profile.
    \item \codeIn{workflow\_id} uniquely identifies a session, allowing the serving layer to track the ongoing lineage and contextual history of a specific execution instance.
    \item \codeIn{agent\_id} uniquely identifies the current node or role in the workflow (\eg, \codeIn{engineer}), explicitly signaling which sub-task is currently executing.
\end{enumerate}

We restrict the API extension to these three identifiers rather than requiring applications to pass explicitly defined dependency graphs or scheduling hints. Crucially, demanding a full workflow graph and agent characteristics upfront is impractical because they are often dynamically determined at runtime. Conversely, requiring application developers to manually tag requests with system-level hints (\eg, ``cache this prefix'') violates the separation of concerns, burdening them with infrastructure resource management. By passing only the identity of workflows and agents, \sysname can easily correlate the request with historical patterns, leaving the burden of prediction and resource allocation entirely to the underlying serving system.

This minimal abstraction enables easy deployments. For framework developers building orchestration layers like LangGraph or AutoGen, integrating with \sysname requires no fundamental architectural changes. They simply need to configure a global API interceptor that automatically attaches the current session and agent variables to the \codeIn{extra\_body} payload before dispatch. Hence, developers building atop these frameworks reap the benefits of advanced system-level optimizations completely transparently, without needing to alter their application logic or write hardware-aware code.

\begin{lstlisting}[language=json, moredelim={[is][\color{blue}]{|b|}{|b|}}, moredelim={[is][\color{red}]{|t|}{|t|}}, caption={API payload: the framework injects lightweight identity metadata via \codeIn{app\_metadata}; our Workflow Profiler intercepts this and injects actionable predictions via \codeIn{sys\_annotations} inside the same \codeIn{extra\_body} block before routing.}, label={lst:api_extension}, float=t]
{
  "model": "meta-llama/Llama-3.1-8B-Instruct",
  "messages": [{"role": "user", "content": "..."}],
  "extra_body": {
    |b|"app_metadata": {
      "workflow_type_id": "coding_assistant",
      "workflow_id": "<session_id>",
      "agent_id": "engineer"
    },|b|
    |t|// Injected internally by the Workflow Profiler
    "sys_annotations": {
      "predicted_output_len": [1000, 1300],
      "predicted_path_regex": "planner -> (explorer)^{||3,4} -> (engineer)^{3,6} -> reviewer -> (engineer^{2-4} -> reviewer)? -> verifier -> terminal",
      "prompt_composition": {"engineer": "You are a helpful engineer. Base code: ${req_12:request:[0,250]} Previous output: ${req_12:response:[0,1024]}"}
    }|t|
  }
}
\end{lstlisting}

\MyPara{Execution trace modeling via regular expressions.} 
Once these annotated requests flow into the system, the workflow profiler asynchronously ingests the logs to reconstruct the application's global behavior. Drawing inspiration from dynamic program analysis~\cite{mining-specifications} and program synthesis~\cite{synthesize-regular-expressions}, we treat the sequence of agent invocations as an execution trace over an alphabet of agent roles. The profiler mines historical traces to construct a \emph{Probabilistic Finite Automaton}~(PFA).

The profiler does not merely predict the single next agent; it synthesizes the PFA into a \emph{regular expression}~(regex) that projects the entire anticipated execution graph. Consider a complex coding workflow: a \codeIn{planner} delegates to several parallel \codeIn{explorers}, followed by an \codeIn{engineer} that iteratively refines code, a \codeIn{reviewer}, and an optional rework loop. Rather than representing this as an infinite, unpredictable cycle\textemdash such as \codeIn{(engineer | reviewer)$^\ast$}\textemdash the profiler bounds the execution to its highly probable structural limits. As shown in the \codeIn{predicted\_path\_regex} field in Listing~\ref{lst:api_extension}, the system captures the parallel fan-out (\codeIn{(explorer)\textsuperscript{||3,4}}), the bounded sequential self-reflection (\codeIn{(engineer)\textsuperscript{3,6}}), and the optional single-retry backward edge (\codeIn{(...)?}).

Furthermore, the profiler models the upcoming prompt structure as a deterministic template. Each placeholder within the \codeIn{prompt\_composition} acts as a strict memory pointer, consisting of three exact parameters: (1) the historical \codeIn{request\_id} of a dependent agent (\eg, \codeIn{req\_1}), (2) a flag indicating whether to extract from that agent's input \codeIn{request} or generative \codeIn{response}, and (3) the specific token index \codeIn{range} to retrieve. Listing~\ref{lst:api_extension} demonstrates how this template captures an agent stitching together the components.

\MyPara{Optimizing for the dominant percentile.} 
A key insight in applying program analysis to agentic workflows is that modeling every possible execution edge case is counterproductive. Due to the inherent stochasticity of LLMs, workflows may occasionally exhibit anomalous transitions, enter chaotic error-recovery loops, or trigger runaway token generation. Attempting to fully account for such outliers leads to infinitely recursive representations and significantly over-provisioned memory reservations.

Instead, \sysname deliberately optimizes for the dominant percentile of executions in both control flow and resource usage. It applies trace filtering to prune low-probability transitions (\eg, those occurring in less than 5\% of cases), yielding a path expression that tightly captures the common execution graph. Likewise, rather than predicting exact output lengths or reserving for the worst-case context window, the profiler derives high-confidence intervals (\eg, the 99th percentile of historical token usage) for each agent. This filtered, probabilistic view aligns with real-world deployments: it reflects the structured routing and guardrails that keep LLMs on stable paths, while bounding memory usage without overreacting to rare, long-tail behaviors.

\MyPara{Actionable request annotation.} 
At inference time, the profiler operates at the gateway. When a request arrives, it extracts the \codeIn{app\_metadata} and queries the filtered models to retrieve the expected output length, resolve the regular expression describing the upcoming workflow graph, and bound the template for the next prompt. It then injects these insights into the \codeIn{sys\_annotations} field within the \codeIn{extra\_body} payload. This single interception step transforms an otherwise opaque request into a fully informed data structure, equipping downstream components with precise execution plans and memory hints for proactive caching, global scheduling, and cluster autoscaling.

\mysubsection{Consuming Predictive Information}
\subsubsection{Speculative Cache Management}
\label{sec:design:cache}
Equipped with the workflow graphs and prompt templates generated by the profiler, \sysname uses a speculative cache manager to mitigate the prefix thrashing and cache misses inherent to multi-agent workloads. By exploiting the future execution graphs embedded in a request's \codeIn{sys\_annotations}, our cache manager shifts memory management from reactive guesswork to a proactive, Belady-like strategy. As shown in Algorithm~\ref{alg:cache_manager}, this is achieved through a combination of early cache eviction and forward cache staging.

\begin{algorithm}[t]
\small
\caption{Speculative Cache Management.}
\label{alg:cache_manager}
\begin{algorithmic}[1]
\Require Current request $R_{curr}$, Hierarchical Cache $C$, GPU Compute Stream $S_{bg}$
\State $regex \gets R_{curr}.\text{predicted\_path\_regex}$
\State $future\_nodes \gets \text{ParseRegex}(regex)$
\State $prompt\_dict \gets R_{curr}.\text{prompt\_composition}$
\State
\Procedure{OnRequestComplete}{} 
    \For{each block $b \in C$}
        \If{$b.\text{lineage} \notin future\_nodes$}
            \State $\text{Free}(b)$ \Comment{Dead tokens, drop immediately}
        \ElsIf{$b.\text{lineage} \in future\_nodes$}
            \State $\text{RetainInHierarchy}(b)$ 
        \EndIf
    \EndFor
\EndProcedure
\State
\Procedure{OnPrefetchRequested}{}
    \For{each $(agent, template) \in prompt\_dict$}
        \State \emph{//$prompt\_dict$ only has one entry}
        \State $prompt_{next} \gets \text{Assemble}(template, \text{History}())$
        \If{$\text{ExistsInCache}(C, prompt_{next})$}
            \State $\text{PromoteToHost}(prompt_{next})$ 
        \ElsIf{$\text{IsIdle}(\text{GPU})$}
            \State $\text{Enqueue}(S_{bg}, \text{ForwardPass}(prompt_{next}))$
        \EndIf
    \EndFor
\EndProcedure
\end{algorithmic}
\end{algorithm}

\MyPara{Early cache eviction~(Lines 5-10).}
At the completion of each request (\codeIn{OnRequestComplete}), the cache manager actively prunes the KV cache to prevent memory bloat and protect critical contexts. By parsing \codeIn{predicted\_path\_regex}, it generates a deterministic set of future agent nodes. It then evaluates the lineage of every block in the cache against this set. Blocks belonging to transient steps that will not be revisited are classified as dead tokens and freed immediately, bypassing the delayed reclamation of standard LRU policies. Conversely, blocks whose lineage appears in the future path are explicitly marked for retention. Rather than simply trapping these retained blocks on the local worker, the manager aggressively writes them out to a shared L3 storage layer accessible by all inference engines. By proactively persisting future-relevant contexts in a global cache space, \sysname decouples cache residency from worker node affinity. This simplifies the cluster's global routing strategy, as subsequent requests in a multi-agent workflow can be scheduled on any available worker without suffering a hard cache miss.

\MyPara{Forward cache staging~(Lines 12-19).}
To mitigate the prefill latency associated with transitioning between agents, the manager proactively prepares the exact context for the anticipated next step (\codeIn{OnPrefetchRequested}). Guided by the \codeIn{prompt\_composition} template provided in the request annotations, the manager assembles the upcoming prompt by injecting the relevant inputs and outputs from the execution \codeIn{History()}. Once the target prompt is assembled, the manager ensures its corresponding prefix cache is staged as close to the compute units as possible. If the cache for this prompt already exists in deeper, shared L3 storage, the manager promotes it to the L2 host DRAM (\codeIn{PromoteToHost}).

Crucially, the system intentionally holds this prefetched context in the L2 host memory rather than immediately pushing it to the highly constrained GPU HBM. This design choice absorbs two practical realities: first, the profiler's prediction of the next agent is speculative and could be incorrect; second, the currently executing request may continue generating tokens for an extended period, and aggressively moving speculative data would prematurely pollute the limited GPU memory. This forward staging acts similarly to logistics positioning: it keeps the necessary memory blocks just one hop away in a less constrained environment, ready for an immediate, on-demand PCIe transfer when the actual request arrives.

However, if the assembled prompt does not exist anywhere in the cache hierarchy, the manager attempts to generate it from scratch. The cache manager opportunistically capitalizes on the GPU's compute units; if it detects idle GPU cycles, it enqueues an asynchronous forward pass to pre-compute the prefix cache in the background. Through this combination of host-level staging and background pre-prefilling, the system maximizes the probability that the prefix is warmed before the framework ever issues the subsequent inference request.

\mysubsubsection{Lookahead Request Scheduling}
\label{sec:design:schedule}

With multi-agent request metadata extracted by the profiler and contexts managed by the cache manager, \sysname optimizes execution across the cluster through a two-tiered scheduling architecture: a statistical capacity-based router at the cluster level, and a graph-driven local scheduler at the worker level.

\begin{algorithm}[t]
\caption{Statistical Capacity-Based Routing.}
\label{alg:routing}
\small
\begin{algorithmic}[1]
\Require Request $R$, Candidate node $N$, Target OOM Threshold $\epsilon$
\State
\Procedure{lRoute}{$R, N, \epsilon$}
    \State $N_{capable} \gets \emptyset$, $target \gets \text{null}$, $max\_headroom \gets 0$
    \For{each $n \in N$}
        \State $P_{oom} \gets \text{Prob}(\sum_{r \in n.\text{active} \cup \{R\}} r.\text{len} > n.\text{capacity})$
        \If{$P_{oom} \le \epsilon$}
            \State $N_{capable} \gets N_{capable} \cup \{n\}$
            \State $headroom \gets n.\text{capacity} - \text{ExpectedSize}(n)$
            \If{$headroom > max\_headroom$}
                \State $max\_headroom \gets headroom$
                \State $target \gets n$
            \EndIf
        \EndIf
    \EndFor
    \State \Return $target$ \Comment{Cache affinity as a tie-breaker}
\EndProcedure
\end{algorithmic}
\end{algorithm}

\MyPara{Statistical capacity routing.} 
At the cluster ingress, the global router assigns incoming requests to specific worker nodes as shown in Algorithm~\ref{alg:routing}. Because LLM output generation is inherently stochastic, relying on static maximum token lengths leads to severe memory underutilization, while relying on simple averages causes request preemptions due to out-of-memory~(OOM). \sysname resolves this by leveraging the high-confidence intervals extracted by the profiler. 

For each request $r_i$, the profiler provides an expected upper-bound length $u_i$ derived from a high-confidence interval~(\eg, the 99th percentile, yielding an error probability $\alpha_i = 0.01$). This guarantees that the probability of the actual length $L_i$ exceeding this bound is $P(L_i > u_i) \le \alpha_i$. To avoid assuming any underlying statistical distribution, \sysname relies on a distribution-free strict capacity reservation. A candidate node with total KV capacity $C$ is considered structurally capable if the sum of the confidence bounds for all active requests $A$ and the new request $R$ fits in memory:
$$ \sum_{r_i \in A \cup \{R\}} u_i \le C $$
When this reservation holds, an OOM event can only occur if the actual token generation violates these individual bounds. By applying the union bound~(Boole's inequality), the router calculates the worst-case joint probability of an OOM event, $P_{oom}$, which is strictly bounded by the sum of the individual error probabilities:
$$ P_{oom} \le \sum_{r_i \in A \cup \{R\}} P(L_i > u_i) \le \sum_{r_i \in A \cup \{R\}} \alpha_i $$
The router enforces that this distribution-free failure probability remains below a strict, user-defined safety threshold ($P_{oom} \le \epsilon$). Among the mathematically safe nodes, it routes the request to the one with the maximum expected headroom ($C - \sum u_i$). While not strictly enforced in the primary capacity bounds, the router uses forward cache staging~(\S\ref{sec:design:cache}) as a deterministic tie-breaker. If multiple nodes offer safe capacity profiles, the router preferentially selects the node that already holds the required L2 host cache, gracefully marrying strict statistical load balancing with instantaneous cache hits.

\begin{algorithm}[t]
\caption{Graph-Driven Global Priority Assignment.}
\label{alg:worker_schedule}
\small
\begin{algorithmic}[1]
\Require Cluster Replica Queues $R_{queues}$, Request $r$
\State
\Procedure{DownstreamIdleRisk}{$R_{queues}, r$}
    \State $risk\_score \gets 0$
    \State $future\_agents \gets \text{GetFutureAgents}(r.\text{regex})$
    \For{each $a \in future\_agents$}
        \If{$R_{queues}[a.model] < \text{IDLE\_THRESHOLD}$}
            \State $E[D_a] \gets \text{ExpectedDist}(\text{r.agent}, a)$
            \State $risk\_score \gets risk\_score + \left(\frac{1}{E[D_m]}\right)$
        \EndIf
    \EndFor
    \State \Return $risk\_score$
\EndProcedure
\State
\Procedure{SetPriority}{$r, R_{queues}$}
    \State $E[D_{remain}] \gets \text{ExpectedRemainingDistance}(r.\text{regex})$
    \State $S_{completion} \gets \frac{1}{E[D_{remain}]}$ 
    \State $S_{unblock} \gets \text{DownstreamIdleRisk}(R_{queues}, r)$
    \State $r.\text{base\_priority} \gets (\omega_1 \times S_{completion}) + (\omega_2 \times S_{unblock})$
\EndProcedure
\end{algorithmic}
\end{algorithm}

\MyPara{Graph-driven global priority assignment.} 
In a multi-agent environment where multiple agents share the same underlying model replicas, standard FCFS scheduling frequently causes cluster-wide starvation. It ignores workflow dependencies, allowing newly spawned, long-running agents to block final-stage agents, which severely delays final-stage completion. To resolve this, \sysname implements a global arbitration strategy~(\codeIn{SetPriority} in Algorithm~\ref{alg:worker_schedule}). Before a request is routed to a worker, the global scheduler assigns it a static \codeIn{base\_priority} score. 

This score explicitly balances two cluster-wide execution goals. First, it accelerates job completion by prioritizing requests closest to the end of their workflow. Because workflow paths are defined by probabilistic regular expressions rather than deterministic graphs, the scheduler calculates the mathematical expectation of the remaining distance to the terminal node ($E[D_{remain}]$). The priority score scales inversely with this expectation, ensuring that agents statistically closer to finishing~(\eg, a final verifier) inherently preempt early-stage agents to quickly flush completed jobs from the system. 

Second, the scheduler actively monitors the real-time queue depths of all model replicas across the cluster~(\codeIn{DownstreamIdleRisk}). If a model replica serving future agents is at risk of becoming idle, the priority of any upstream request that will eventually unblock it is boosted. Crucially, this boost is weighted inversely by the expected distance between the current agent and the starved model ($E[D_m]$). A request that is only one step away from feeding an idle GPU receives a higher priority spike, whereas a request ten steps away receives a negligible boost. This ensures that no specialized inference worker sits idle while waiting for a stalled upstream dependency.

\noindent\textbf{\textit{Periodical local batch creation.}}
Once routed to a worker, execution is governed by iteration-level dispatching. At the start of every scheduling window, the local worker dynamically recalculates the effective priority of all requests by adding an aging factor\textemdash scaled by accumulated wait time\textemdash to the global \codeIn{base\_priority}. The worker then sorts the pool and dispatches the highest-priority subset that fits within memory bounds. This ensures early-stage agents are mathematically protected from permanent starvation while preserving the global graph's intended execution order.

\noindent\textbf{\textit{Priority-aware preemption recovery.}} 
When tail-latency generations occasionally force a worker into memory exhaustion, \sysname overrides standard FCFS eviction policies. Instead of blindly appending evicted requests to the back of the queue, the local scheduler specifically targets the active request with the lowest dynamically calculated priority~(\eg, an early-stage agent) as the preemption victim. This request is paused and re-inserted into the local queue. Because the queue is strictly re-sorted before every iteration, critical-path executions remain completely unperturbed by localized memory spikes, leading to high effective resource utilization and minimized workflow-level waiting time.

\mysubsubsection{Phase-Adaptive Autoscaling}
\label{sec:design:scale}
While lookahead request scheduling~(\S\ref{sec:design:schedule}) mitigates load imbalance, it cannot create new capacity. Agentic applications frequently exhibit extreme burstiness driven by synchronized triggers~(\eg, cron jobs) or massive structural fan-outs. During these dramatic shifts, scheduling alone cannot prevent queue saturation; the cluster must physically scale its model replicas.

Current serverless LLM solutions rely on reactive autoscaling, which fails during the aggressive, instantaneous phase shifts typical of multi-agent execution. Because loading massive model weights into GPU memory takes tens of seconds, reactive provisioning is not merely too slow\textemdash it is actively detrimental. In a constrained cluster, spinning up a new model often forces the blind eviction of existing ones. By the time the new replica is finally ready, the transient burst has often subsided, leaving the system with an idle model and a destroyed cache of other essential models, triggering a destructive cascade of cold starts. 

\sysname overcomes this by leveraging the regular path expressions extracted by the workflow profiler to introduce a predictive control plane. Instead of reacting to queue buildup, it forecasts imminent bursts and proactively provisions replicas before the traffic shift occurs, entirely avoiding the reactive resource-thrashing cycle.

\begin{algorithm}[t]
\small
\caption{Phase-Adaptive Autoscaling.}
\label{alg:autoscaler}
\begin{algorithmic}[1]
\Require Active Requests $A$, Current Replicas $R_{curr}$, Look-Ahead Horizon $H$
\State
\Procedure{EstimateImminentDemand}{$A, H$}
    \State $D \gets \text{InitializeZeroMap}()$
    \For{each $r \in A$}
        \State $imminent\_agents \gets \text{ProjectGraph}(r.\text{regex}, H)$
        \For{each $a \in imminent\_agents$}
            \State $D[a.\text{model}] $+=$ \text{EstimatedLoad}(a)$
        \EndFor
    \EndFor
    \State \Return $D$
\EndProcedure
\State
\Procedure{AutoscaleCluster}{}
    \State $D \gets \text{EstimateImminentDemand}(A, H)$
    \State $R' \gets \text{InitializeMap}()$
    \For{each $m \in \text{RegisteredModels}()$}
        \State $R'[m] \gets \text{EstimateReplicas}(D[m])$
    \EndFor
    \For{each $m \in \text{RegisteredModels}()$}
        \If{$R'[m] < R[m]$}
            \State $\text{ScaleDownOnIdle}(m, R[m] - R'[m])$
        \EndIf
    \EndFor
    \For{each $m \in \text{RegisteredModels}()$}
        \If{$R'[m] > R[m]$}
            \State $\text{ScaleUp}(m, R'[m] - R[m])$
        \EndIf
    \EndFor
\EndProcedure
\end{algorithmic}
\end{algorithm}

\MyPara{Proactive scale-up for structural shifts.} 
Rather than waiting for queues to spike, \sysname continuously evaluates the state of all active requests against their workflow graph to forecast imminent demand~(\codeIn{EstimateImminentDemand} in Algorithm~\ref{alg:autoscaler}). By projecting the execution graph forward by a defined step horizon~($H$), the autoscaler calculates which models will be extensively utilized in the immediate future. 

For example, if the cluster currently has fifty active requests executing in a \codeIn{planner} phase, and the underlying regex dictates a transition of \codeIn{planner $\rightarrow$ (explorer)\textsuperscript{||10}}, the autoscaler mathematically knows that 500 requests for the \codeIn{explorer} model are arriving in the near future. Before the \codeIn{planner} agents even finish generating their outputs, \sysname proactively triggers the background provisioning of additional \codeIn{explorer} model replicas. By the time the framework actually dispatches the massive fan-out burst, the new GPU nodes are already warmed and ready, effectively alleviating burst-induced queuing and hiding cold-start latencies.

\MyPara{Predictive scale-down and resource reclamation.} 
In a resource-constrained cluster, scaling up one specialized model often requires evicting another. Standard systems rely on static keep-alive timeouts~(\eg, waiting 5 minutes before spinning down an idle model), which hoards precious GPU memory during critical phase shifts. \sysname accelerates this reclamation process through predictive scale-down.

Because the autoscaler possesses the full structural map of the active workloads, it can identify which models will \emph{not} be called in the near future. This includes models powering agents that are strictly in the past~(\eg, the \codeIn{planner} once all workflows have transitioned to the \codeIn{explorer} phase) and models powering agents that are topologically too far away to warrant immediate warming~(\eg, a \codeIn{final\_verifier} that is many steps away from the current execution frontier). To maximize the opportunity for a safe scale-down, the autoscaler coordinates directly with the global scheduler. Once a model is identified as obsolete for the immediate look-ahead horizon, the autoscaler signals the scheduler to cease load-balancing new requests to those specific replicas. This active routing exclusion allows the replicas to drain rapidly. As soon as a targeted replica's immediate queue is empty, \sysname forcefully deprovisions it without waiting for an arbitrary timeout. This aggressive, graph-driven reclamation ensures that cluster resources are continuously freed and reallocated to support the models required for the immediate execution phase.

\section{Discussion}
\label{sec:discuss}

\MyPara{Support for many distinct agentic workflows.} Currently, \sysname is primarily designed for environments running a small, targeted set of agentic applications. When scaling the system to concurrently serve a massive variety of distinct workflows, \sysname successfully maintains its core advantage of minimizing the global average JCT. However, in such a highly heterogeneous environment, relying on our current, relatively simple scheduling heuristics may inadvertently sacrifice the performance of certain individual workflows to maximize overall throughput. To effectively balance cluster-wide resource efficiency with other user requirements~(\eg, per-workflow SLOs or fairness guarantees), \sysname's lookahead scheduler can be naturally extended. By directly incorporating per-workflow priority weights and explicit latency goals into our scheduling algorithm, the system can make more nuanced, SLO-aware allocation decisions that protect mission-critical workflows while still preserving aggregate optimizations.

\MyPara{Cold starts and workflow drift.}
\sysname's reliance on historical traces introduces challenges during initial deployment~(cold starts) and application updates~(workflow drift). To prevent insufficient or stale data from triggering unsafe routing, \sysname can pair explicit API versioning with a ``shadow profiling'' phase. When a new workflow version is detected via \codeIn{app\_metadata}, the gateway temporarily routes its traffic through a standard, reactive fallback while profiling it in the background. Once safe statistical confidence intervals are established, the system seamlessly promotes the workflow to the proactive execution tier, ensuring stability across lifecycle transitions.

\MyPara{Open-ended and adversarial workflows.}
\sysname relies on the structural predictability typical of enterprise applications, which apply routing guardrails and assume trusted inputs. In fully unconstrained or adversarial scenarios, execution graphs devolve into unpredictable meshes, rendering our proactive optimizations ineffective. To accommodate such environments, future deployments could adopt a hybrid architecture: \sysname would serve verified, predictable workflows, while offloading highly deviant or adversarial requests to an isolated, reactively scheduled fallback environment.
\mysection{Implementation}
\label{sec:impl}

\begin{figure*}[t]
  \centering
  \includegraphics[width=0.9\linewidth]{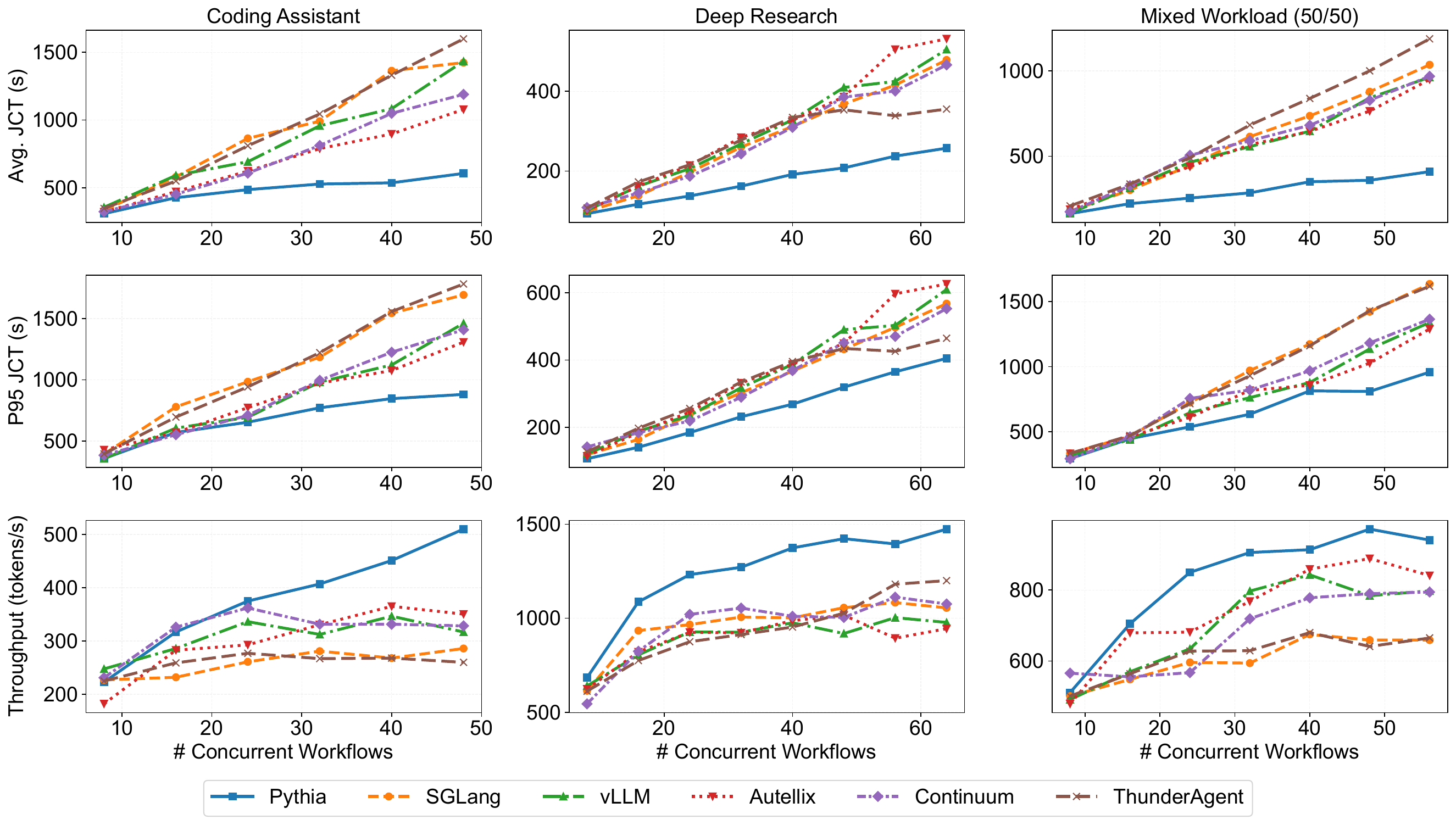}
  \vspace{-1.5em}
  \caption{End-to-end experiments.}
  \label{fig:e2e}\vspace{-1em}
\end{figure*}

We implemented \sysname on top of SGLang. The Workflow Profiler and Global Request Scheduler replace the default SGLang router, intercepting standard OpenAI-compatible requests to extract \codeIn{app\_metadata}, injecting \codeIn{sys\_annotations}, and evaluating the statistical capacity bounds~(\S\ref{sec:design:schedule}) before routing. At the engine level, we integrated the Speculative Cache Manager~(\S\ref{sec:design:cache}) into SGLang's Hierarchical Cache. Furthermore, we replaced SGLang's default scheduler with our graph-driven iteration dispatcher, ensuring that requests are strictly sorted by their dynamic \codeIn{priority} score at the start of every iteration. Finally, the Phase-Adaptive Autoscaler~(\S\ref{sec:design:scale}) runs as a periodic daemon on the control plane, provisioning and deprovisioning SGLang model replicas across the GPU fleet based on the projected demand.
\mysection{Evaluation}
\label{sec:evaluation}

\MyPara{Testbed.} We have evaluated \sysname on a server equipped with two AMD EPYC 7J13 CPUs, 1.7TB of DRAM, and eight NVIDIA A100 GPUs~(80GB HBM each) interconnected via NVLink.

\MyPara{Baselines.} We compare \sysname with a comprehensive suite of state-of-the-art LLM serving systems and agent-aware frameworks:

\squishlist
\item \textbf{vLLM~\cite{vLLM:SOSP23}:} vLLM~(v0.17.1) deployed with vLLM Production Stack~\cite{vllm-prod-stack}~(v0.1.10) router.
\item \textbf{SGLang~\cite{SGLang:NIPS24}:} SGLang~(v0.5.9) utilized alongside the gateway.
\item \textbf{Autellix~\cite{Autellix}:} A specialized agentic LLM serving system that provides integrated scheduling optimizations across the router and inference engine. Because Autellix is not open-sourced, we implemented the same routing and scheduling policy on top of vLLM~(v0.17.1).
\item \textbf{Continuum~\cite{Continuum}:} A single-node serving system optimized for ReAct~\cite{ReAct:ICLR23} agents; we use it with the vLLM Production Stack router~(v0.1.10) for fair comparison. We only allow agents that can potentially have cache hits to pin their cache.
\item \textbf{ThunderAgent~\cite{ThunderAgent}:} A program-aware global LLM request scheduler for agentic applications, which we use SGLang~(v0.5.9) as the underlying inference engine. Because ThunderAgent is initially designed for a single model, we changed its single-queue design to one queue per model. We also changed its scheduling interval from the default 5 seconds to 0.5 seconds to reduce unnecessary queuing delay.
\squishend

All systems use Mooncake~\cite{Mooncake:FAST25} for the L3 prefix cache, while vLLM-based baselines are integrated via LMCache~\cite{LMCache}.

\MyPara{Workloads.} We evaluate \sysname using two representative agentic workflows ported to the AutoGen~\cite{Autogen:COLM24} framework: a proprietary internal coding assistant and a Deep Research Agent~\cite{Salesforce}. Using these applications, we execute two demanding benchmarks: \textbf{SWE-bench Pro}~\cite{SWEBenchPro}, a rigorous coding benchmark mirroring complex production patterns, and \textbf{Deep Research Bench}~\cite{DeepResearchBench}, designed to evaluate PhD-level research tasks.

Our evaluation utilizes real-world traces collected from these applications across a wide portfolio of frontier models\textemdash including those from Anthropic, OpenAI, and Google\textemdash assigned to various agent roles. To ensure high-fidelity simulation of agentic behavior, our trace collection and evaluation involve live integration with real execution environments and production-grade search APIs.

To facilitate deployment on our testbed, we map these frontier models to smaller-scale models~(3B--14B) from the Qwen and Llama families, employing tensor parallelism~(TP) degrees ranging from one to four.

\subsection{End-to-End Results}

We first evaluated \sysname against all five baseline systems under different numbers of concurrent workflows. We compare average JCT, P95 JCT, and throughput~(output tokens generated per second).

\MyPara{Average JCT.}
Across all workloads, \sysname reduces average JCT by 1.38--2.9$\times$ at peak concurrency. Under low load, gains stem primarily from our speculative cache management hiding prefill latency. As density increases, \sysname's graph-driven global arbitration prevents the catastrophic HoL blocking that plagues baselines\textemdash particularly in heterogeneous mixed workloads. While partially agent-aware systems~(Continuum, Autellix) generally outperform agnostic engines, and ThunderAgent shows some resilience under heavy Deep Research loads, \sysname uniquely maintains steady performance across all scenarios.

\begin{figure}[t]
  \centering
  \begin{minipage}[t]{0.45\columnwidth}
    \centering
    \includegraphics[width=\linewidth]{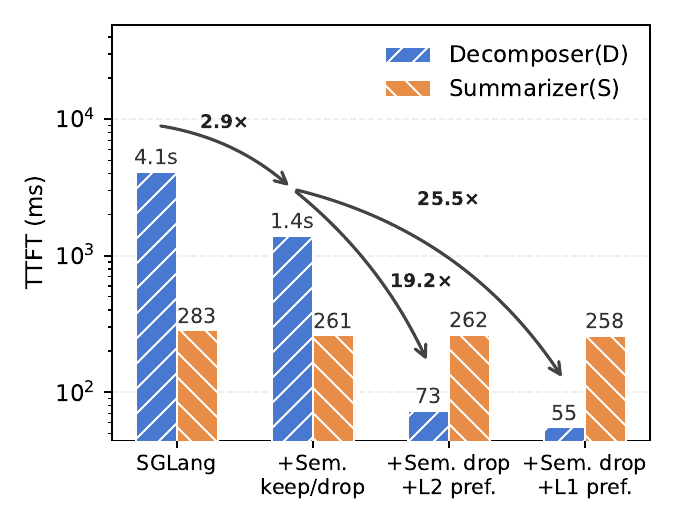}
    \par\vspace{0.5pt}
    \small (a) \sysname's semantic-aware keep/drop and speculative prefetch improve TTFT.
    \label{fig:hicache-ttft}
  \end{minipage}\hfill
  \begin{minipage}[t]{0.49\columnwidth}
    \centering
    \includegraphics[width=\linewidth]{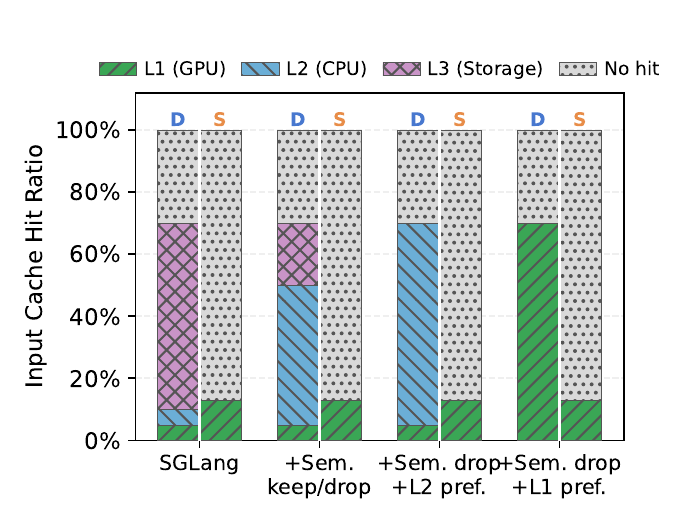}
    \par\vspace{0.5pt}
    \small (b) Input cache hit ratio broken down by tier.
    \label{fig:hicache-cache-hit}
  \end{minipage}
  \caption{Workflow-aware speculative prefix cache management improves TTFT latency and prefix cache hit ratio from the lower cache tier.}
  \label{fig:hicache-micro}
\end{figure}

\MyPara{P95 JCT.}
While approximate SRTF scheduling typically degrades tail latency by starving early-stage requests, \sysname defies this trade-off, reducing P95 JCT by 1.15--2.02$\times$ under high concurrency. Instead of simply favoring short outputs, \sysname dynamically boosts upstream requests that unblock downstream models. Paired with priority aging, this holistic strategy prevents complex multi-step workflows from languishing in queues, ensuring highly predictable tail performance even under extreme load.

\MyPara{Throughput.}
\sysname delivers a 1.12--1.96$\times$ throughput improvement across all evaluated workloads. This gain stems from two complementary mechanisms: the statistical capacity router leverages output length predictions to evenly distribute load and minimize localized memory contention, while the graph-driven scheduler prioritizes critical-path tasks to prevent downstream model idleness, ultimately maximizing aggregate cluster utilization.

\mysubsection{Performance Analysis}

\MyPara{Cache efficiency.} We take the deep research workflow as an example to show the effectiveness of \sysname's speculative cache management. A deep research workflow can repeat a few turns where Decomposer shares context across turns but Summarizer receives a large amount of new context from Researchers. As shown in Figure~\ref{fig:hicache-micro}, collaboratively applying proactive keep/drop actions and timely prefetch, TTFT of Decomposer is reduced by 25.5$\times$ with cached tokens reused from L1/L2 cache.

\MyPara{Comparison of different scheduling algorithms.}
\begin{figure}[t]
  \centering
  \includegraphics[width=\columnwidth]{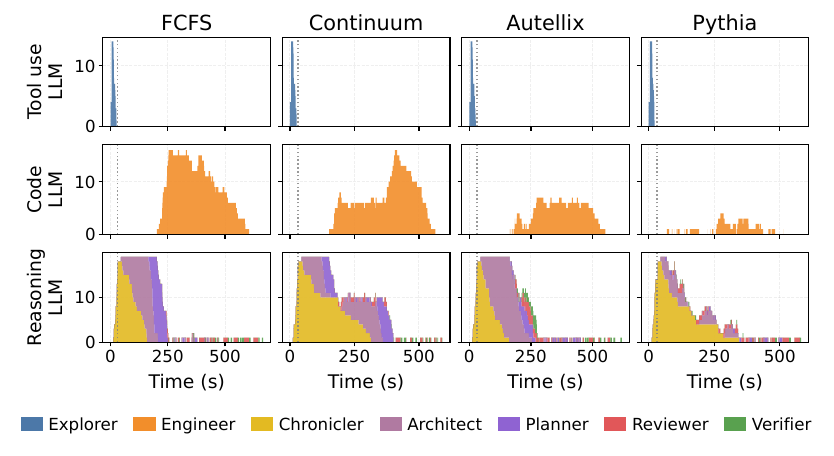}
  \caption{GPU queue dynamics under different scheduling policies for coding assistant. \sysname achieves lower queuing delay and more balanced GPU utilizations.}
  \label{fig:queue-breakdown}
\end{figure}
To analyze the effectiveness of our graph-driven global priority assignment, we break down the queuing dynamics when serving bursts of coding-assistant workflows. Figure~\ref{fig:queue-breakdown} shows that all three baselines build substantially larger queues on both the codeLLM and the shared reasoning LLM. FCFS blindly serves earlier-arriving requests, causing upstream reasoning requests to block later-stage review and verification, which delays completion and eventually creates backlog on the codeLLM. Continuum similarly preserves job-level arrival order across the workflow, slowing the drain of requests that are already near completion. Autellix improves over FCFS, but because it is unaware of workflow position and downstream bottlenecks, it still over-admits upstream requests and leaves significant queue buildup. In contrast, \sysname uses workflow structure and remaining work to prioritize completion-critical requests and throttle upstream injection when bottlenecks emerge, thereby preventing queue waves across GPUs and yielding lower JCT and higher throughput.

\begin{figure}
    \centering
    \includegraphics[width=0.7\linewidth]{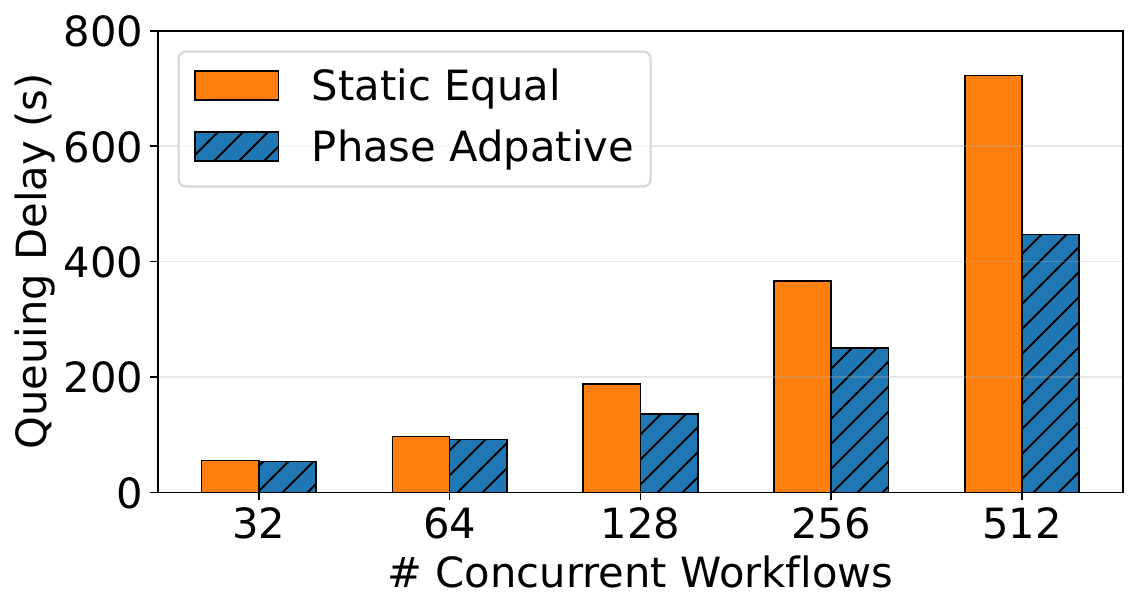}
    \caption{Queuing delay under different scaling strategies.}
    \label{fig:autoscaling}
\end{figure}

\MyPara{Effectiveness of autoscaling.} Figure~\ref{fig:autoscaling} compares \sysname with static resource allocation as we vary the number of concurrent workflows. \sysname reduces queuing delay by 1.49$\times$ relative to static model placement by proactively rebalancing resources via model autoscaling. The key advantage is that \sysname predicts workflow phase load patterns and allocates additional model instances to stages undergoing bursty request arrivals. Static placement, by contrast, cannot respond to these shifting bottlenecks. As concurrency increases, more workflows simultaneously traverse different stages, amplifying burstiness and causing long request queues under fixed provisioning.

\mysubsection{Ablation Studies}

\begin{figure}
    \centering
    \includegraphics[width=0.99\linewidth]{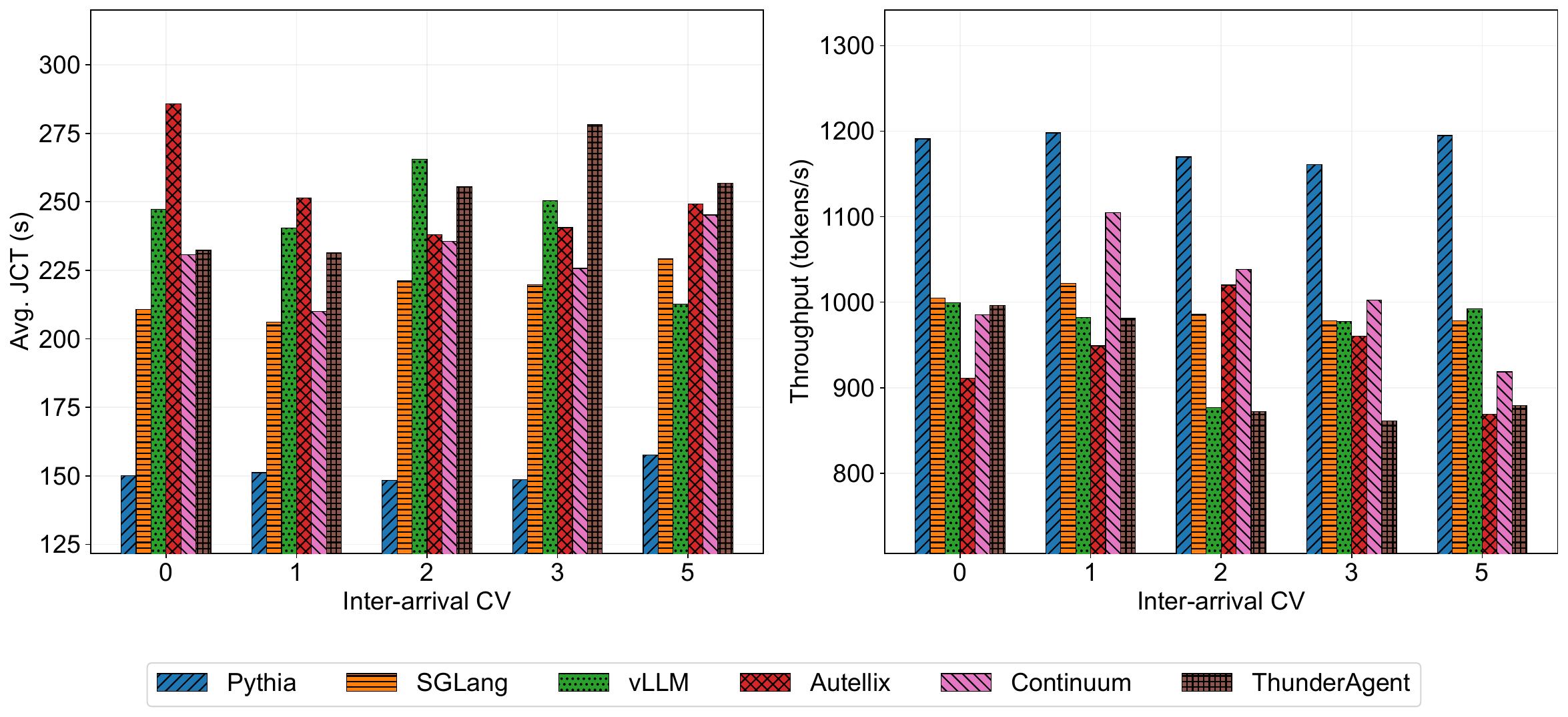}
    \caption{Sensitivity to load burstiness.}
    \label{fig:elastic-engine}
\end{figure}

\MyPara{Sensitivity to load burstiness.} As agentic workflows often arrive in a batch, it's important for the underlying serving system to saturate bursts. In Figure~\ref{fig:elastic-engine}, we show average JCT and throughput under different CVs for arrival interval~(0: stable, 1: Poisson-like, >1: bursty). While \sysname's performance is relatively consistent, turbulence is observed across baselines.

\begin{figure}[t]
  \centering
  \begin{minipage}[t]{0.8\columnwidth}
    \centering
    \includegraphics[width=\linewidth]{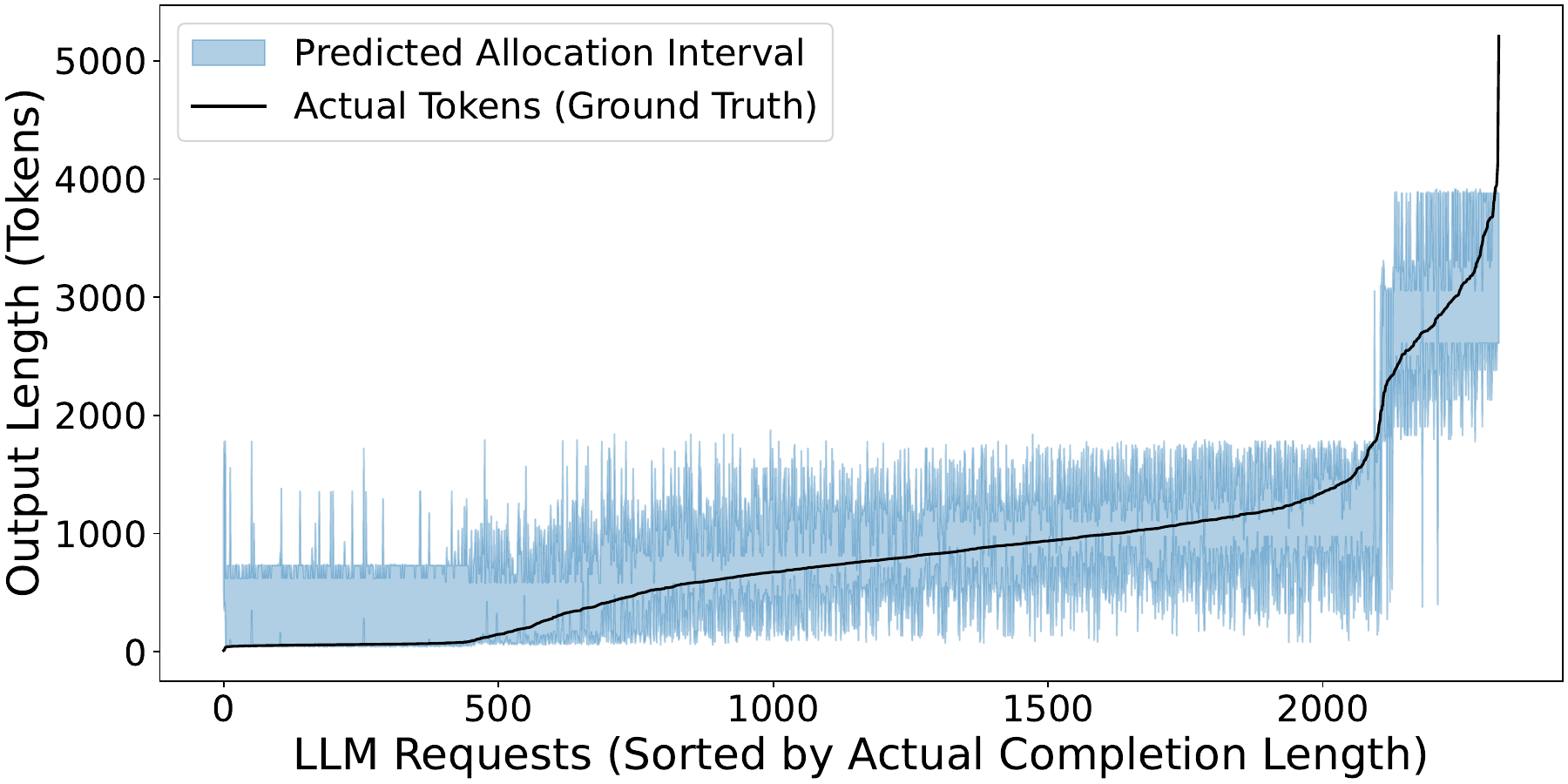}
    \par\vspace{0.5pt}
    \small (a) Coding assistant.
    \label{fig:acc:coding}
  \end{minipage}
  
  \vspace{1em} 
  
  \begin{minipage}[t]{0.8\columnwidth}
    \centering
    \includegraphics[width=\linewidth]{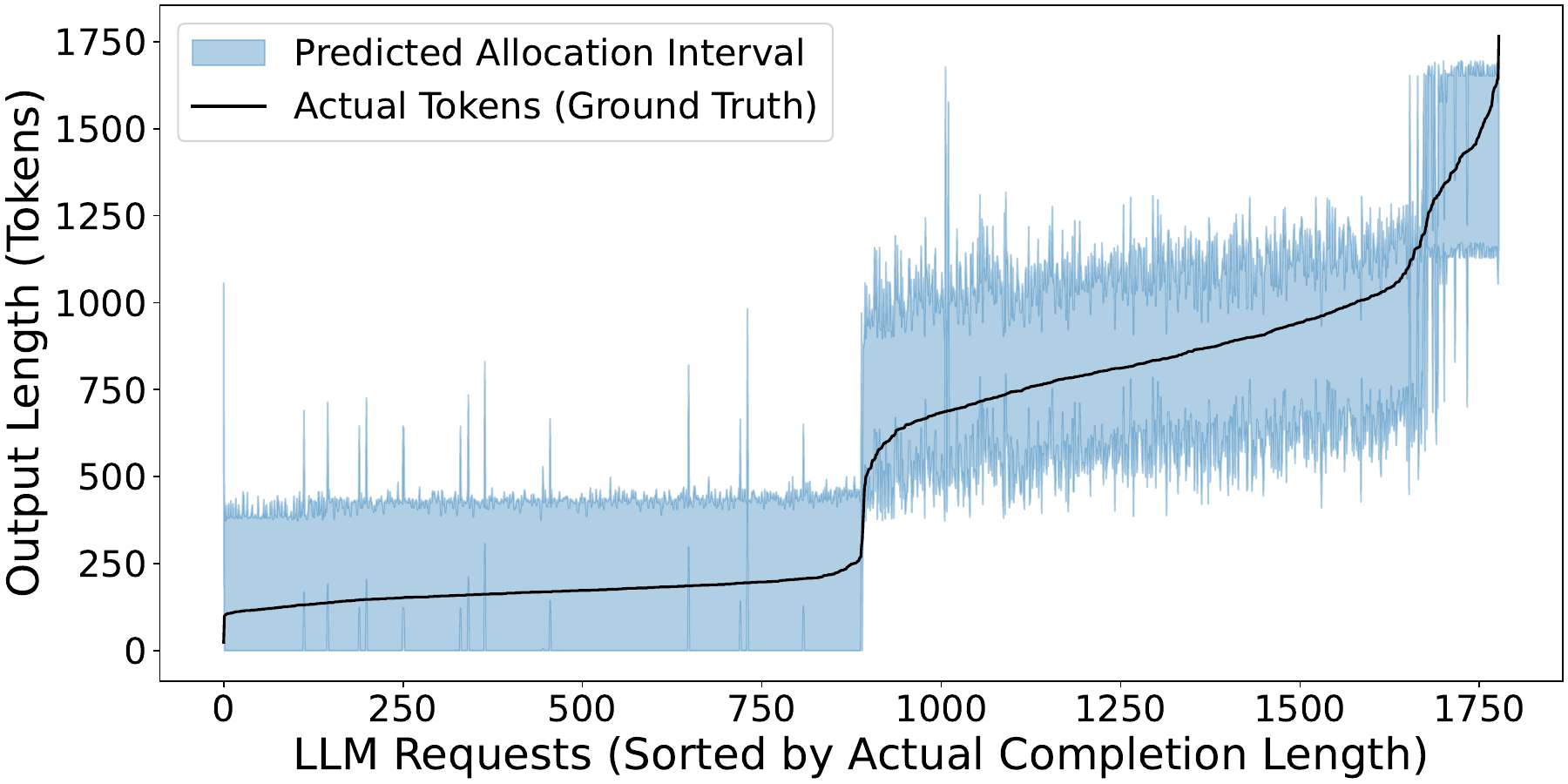}
    \par\vspace{0.5pt}
    \small (b) Deep research.
    \label{fig:acc:dr}
  \end{minipage}
  \caption{Predicted output lengths vs. actual lengths.}
  \label{fig:acc}
\end{figure}

\MyPara{Accuracy of the predictor.} In Figure~\ref{fig:acc}, we show the predicted output length intervals against the ground truth on the test set. For both coding assistant and deep research, \sysname's predictor is able to provide well-bounded predictions across most data points except a few outliers. \sysname can also accurately provide workflow graphs of these applications whose patterns are generally stable.

\section{Related Work}
\label{sec:related}

\MyPara{LLM serving systems.} Modern serving engines are highly optimized across multiple layers. For scheduling, Orca~\cite{Orca:OSDI22} introduces iteration-level scheduling to handle varying request lengths, while FastServe~\cite{FastServe} uses preemption to mitigate HoL blocking. To manage heterogeneous execution stages, systems explore either disaggregating~\cite{DistServe:OSDI24, Splitwise:ISCA24, Dejavu:ICML24, Megascale-Infer:SIGCOMM25}~(\eg, Prefill–Decode disaggregation) or co-locating~\cite{Sarathi-Serve:OSDI24, NanoFlow:OSDI25}~(\eg, Chunked Prefill) these phases. Memory management is equally critical: PagedAttention~\cite{vLLM:SOSP23} uses OS virtual memory abstractions to reduce GPU memory fragmentation, while other works accelerate the prefill stage by exploiting caching and memory hierarchies~\cite{SGLang:NIPS24, LMCache, Mooncake:FAST25, SYMPHONY, Strata}. At the operator level, libraries like FlashInfer~\cite{FlashInfer:MLSys25} and FlashAttention~\cite{FlashAttention:NIPS22} provide the foundational high-performance attention kernels. Finally, multi-LLM serving systems focus on spatial and temporal multiplexing to maximize the density of models co-located on shared hardware~\cite{Prism, Aegaeon:SOSP25, MuxServe:ICML24, ServerlessLLM:OSDI24, HydraServe}. Unlike \sysname, which optimizes for end-to-end agentic workflows, these systems primarily target granular request- and token-level metrics.

\MyPara{System optimizations targeting agentic AI.}
Recent work has recognized that serving agentic AI applications introduces new system challenges beyond traditional LLM inference, motivating dedicated optimizations in specialized runtimes and serving frameworks. Pancanke~\cite{Pancake} designs an efficient multi-tier agent memory system, and DualPath~\cite{DualPath} optimizes the network bottleneck of remote prefix cache for agentic workloads. Systems such as ThunderAgent~\cite{ThunderAgent}, Ayo~\cite{Ayo:ASPLOS25}, and Autellix~\cite{Autellix} focus on improving request orchestration of long-horizon complex agent pipelines, while KVFlow~\cite{KVFlow} and Continuum~\cite{Continuum} propose memory optimizations through cross-step prefix cache sharing in order to reduce redundant prefills. However, these systems do not fully exploit predictable program semantics for fine-grained serving optimizations as \sysname does.

\mysection{Conclusion}
\label{sec:conclusion}

The structured nature of multi-agent LLM workloads fundamentally changes the serving design space. By exposing workflow semantics, \sysname transforms opaque requests into predictable executions, unlocking proactive optimizations impossible in conventional systems. Our evaluation proves that leveraging this predictability is critical to overcoming cache inefficiency, resource imbalance, and burst-induced congestion.

\label{lastpage}

{
\bibliographystyle{ieeetr}
\balance
\bibliography{ref-simple}}


\end{document}